\newif\ifShowKeys
\ShowKeystrue
\ShowKeysfalse

\documentclass[11pt,a4paper]{article} 
\pdfoutput=1
\usepackage[no-natbib-sort]{my-jheppub}
  \ifShowKeys \usepackage[notcite]{showkeys} \fi

\usepackage{amsmath, amssymb}

\usepackage{bm}
\usepackage{environ}
\usepackage{mathrsfs}
\usepackage{array,arydshln}
\usepackage{booktabs}

\usepackage{graphicx,epsfig}
\usepackage{epic}

\usepackage{tensor} 							% Ratcliffe package to write tensors

\usepackage{float}
\usepackage[dvipsnames]{xcolor}
\definecolor{maroon}{rgb}{0.8,0.3,0.}

\usepackage{slashed}
\usepackage[nodayofweek]{date time}

\usepackage{hyperref}

\usepackage{aurical}
\usepackage[T1]{fontenc}

\usepackage{framed}
\definecolor{shadecolor}{RGB}{255, 230, 204}

\allowdisplaybreaks

\newmuskip\pFqmuskip

\newcommand*\pFq[6][8]{%
  \begingroup % only local assignments
  \pFqmuskip=#1mu\relax
  \mathcode`\,=\string"8000
  \begingroup\lccode`\~=`\,
  \lowercase{\endgroup\let~}\pFqcomma
  {}_{#2}F_{#3}{\left[\genfrac..{0pt}{}{#4}{#5};#6\right]}%
  \endgroup
}

\newcommand*\pFtildeq[6][8]{%
  \begingroup % only local assignments
  \pFqmuskip=#1mu\relax
  \mathcode`\,=\string"8000
  \begingroup\lccode`\~=`\,
  \lowercase{\endgroup\let~}\pFqcomma
  {}_{#2}\widetilde{F}_{#3}{\left[\genfrac..{0pt}{}{#4}{#5};#6\right]}%
  \endgroup
}

\newcommand{\pFqcomma}{\mskip\pFqmuskip}

\makeatletter
\DeclareFontFamily{OMX}{MnSymbolE}{}
\DeclareSymbolFont{MnLargeSymbols}{OMX}{MnSymbolE}{m}{n}
\SetSymbolFont{MnLargeSymbols}{bold}{OMX}{MnSymbolE}{b}{n}
\DeclareFontShape{OMX}{MnSymbolE}{m}{n}{
    <-6>  MnSymbolE5
   <6-7>  MnSymbolE6
   <7-8>  MnSymbolE7
   <8-9>  MnSymbolE8
   <9-10> MnSymbolE9
  <10-12> MnSymbolE10
  <12->   MnSymbolE12
}{}
\DeclareFontShape{OMX}{MnSymbolE}{b}{n}{
    <-6>  MnSymbolE-Bold5
   <6-7>  MnSymbolE-Bold6
   <7-8>  MnSymbolE-Bold7
   <8-9>  MnSymbolE-Bold8
   <9-10> MnSymbolE-Bold9
  <10-12> MnSymbolE-Bold10
  <12->   MnSymbolE-Bold12
}{}

\let\llangle\@undefined
\let\rrangle\@undefined
\DeclareMathDelimiter{\llangle}{\mathopen}%
                     {MnLargeSymbols}{'164}{MnLargeSymbols}{'164}
\DeclareMathDelimiter{\rrangle}{\mathclose}%
                     {MnLargeSymbols}{'171}{MnLargeSymbols}{'171}
\makeatother

\newcommand{\be}{\begin{equation}}
\newcommand{\ee}{\end{equation}}

\newcommand{\mc}{\mathcal }

\newcommand{\bv}{\begin{verbatim}}
\newcommand{\ev}{\end{verbatim}}

\newcommand{\la}{\label}

\newcommand{\eps}{\varepsilon}

\newcommand{\muMR}{\mu_{\scriptscriptstyle{{\rm  MR}}}}
\newcommand{\muDR}{\mu_{\scriptscriptstyle{{\rm DR}}}}

\newcommand{\red}[1]{\textcolor{red}{#1}}

\def \ov {\over}\def \te  {\textstyle} 
\def \ci {\cite}
\def \foot {\footnote}
\def \N {{\cal N}}

\def \m {\mu}
\def \n {\nu}
\def \del{\partial}

\newcommand{\rf}[1]{(\ref{#1})}

\def \l {\lambda}
\def\z{\zeta}
\def \iffa {\iffalse}
 \def \no {\notag}\def \OO {{\cal{O}}}
 \def \ha {{\textstyle{1 \ov 2}}}

\title{On non-supersymmetric  generalizations of  the Wilson-Maldacena loops in $\mc N=4$ SYM}

\author[a]{Matteo Beccaria} 
\author[b]{ and \ \ Arkady A. Tseytlin\footnote{Also at Lebedev Institute, Moscow.}} 

\abstract{
Building on our previous work  arXiv:1712.06874 we  consider  one-parameter 
 Polchinski-Sully  generalization  of  the  Wilson-Maldacena (WM)   loops in 
 planar $\N=4$ SYM theory. This breaks  local supersymmetry of WM loop 
 and leads to running of the deformation parameter $\z$. 
 We compute the three-loop   ladder diagram  contribution to the expectation value 
 of the circular loop  which gives the full answer   for  large  $\z$. 
 The  limit  $\z\gg1, \,\lambda\,\z^2=$ fixed in which   the expectation value is determined 
 by the  Gaussian  adjoint scalar  path integral 
  might be exactly solvable despite the lack of global supersymmetry. 
  We  study    similar   generalization of the $\frac{1}{4}$-BPS 
  "latitude"  WM  loop   which depends on two parameters  (in addition to
  the  't Hooft  coupling $\l$). 
One may also introduce another   supersymmetry-breaking  parameter --
the winding number  of the  scalar coupling  circle. 
We   find  the two-loop expression for the  expectation  value 
of the associated  loop by  combining 
the ladder diagram  contribution with an indirect determination
 of the non-ladder 
contribution using 1d  defect CFT perturbation theory.
}

\affiliation[a]{Dipartimento di Matematica e Fisica Ennio De Giorgi,\\
Universit\`a del Salento \& INFN, Via Arnesano, 73100 Lecce, 
Italy} 

\affiliation[b]{The Blackett Laboratory, Imperial College, London SW7 2AZ, U.K.}
                       
\emailAdd{matteo.beccaria@le.infn.it} \emailAdd{tseytlin@imperial.ac.uk}

\begin{document}
\date{\currenttime}
%\begin{flushright}\boxed{\small{\tt \today \ \ - \ \  \currenttime }}\end{flushright}
\begin{flushright}\small{  Imperial-TP-AT-2018-{01}}\end{flushright}				% report number
\maketitle

%\newpage

\section{Introduction}
\def \const {{\rm const} }

As proposed in \cite{Polchinski:2011im} and recently discussed in \cite{Beccaria:2017rbe}, 
it is interesting to consider a  one-parameter   family of Wilson loop operators defined in 
$\mc N=4$ SYM  by  %($n^m n^m =1$) 
\be
\la{1.1}
W^{(\zeta)}(C) = \frac{1}{N}\,\text{Tr}\,\mc P\,\exp\oint_{C}d\tau\,\big[
i\,A_{\mu}(x)\,\dot x^{\mu}+\z\,|\dot x|\,\Phi_{m}(x) \,n^{m}\big] \ .
\ee
Here $n_m(\tau)$  is a unit  6-vector  and $\z$  is a real parameter. 
%where $\z$ is a constant parameter (in general depending on RG scale)  and . 
%If $C$ is a circle and  unit vector $n^m$ is  independent of $\tau$ (i.e. represents a  point of $S^{5}$),
 The operator $W^{(\zeta)}(C)$
interpolates between the standard Wilson loop $W^{(0)}$ and the %$\frac{1}{2}$-BPS
``locally-supersymmetric"  Wilson-Maldacena  (WM) 
loop $W^{(1)}$ \cite{Maldacena:1998im, Rey:1998ik}.
In particular,  one may  consider the case of a circular  loop    with $n^m=\const $
for which \rf{1.1}   with  $\z=1$ is  $\frac{1}{2}$-BPS (preserves 8 out of the 8+8 superconformal symmetries
of $\mc N=4$ SYM)   but  has no global supersymmetry if $\z\not=1$.

The scale dependence of the (renormalized) coupling $\z$ is controlled at one-loop by the 
beta function  \cite{Polchinski:2011im}
\be
\la{1.2}
\beta(\z) = \mu { d \z \ov d \mu} = \frac{\lambda}{8\,\pi^{2}}\,\z\,(\z^{2}-1)+\mc O(\lambda^{2}),
\ee
where $\lambda = g^{2}\,N$ is the  't Hooft coupling and we consider  the planar limit. 
 $\z=0,1$
are expected to be conformal points to  all orders in $\lambda$. 

In the case of the circular loop with constant $n^m$ 
the two-loop weak-coupling expression for  $\langle W^{(\z)}\rangle$  is  found to be \cite{Beccaria:2017rbe}
\be
\la{1.3}
\langle W^{(\z)}\rangle = 1+\frac{\lambda}{8}+\lambda^{2}\,\Big[
\frac{1}{192}+\frac{1}{128\,\pi^{2}}\,(\z^{2}-1)^{2}\Big]+\mc O(\lambda^{3}).
\ee
 Starting at $\l^3$ order, the   UV divergences in \rf{1.3}  do not cancel
   but can be absorbed into a   renormalization of $\z$. 
 
As discussed in \cite{Beccaria:2017rbe}, the expectation value $\langle W^{(\z)}\rangle$ may 
be interpreted as the partition function of an effective  ``defect''
 1d QFT which   becomes  conformal at $\z=0,1$. 
Expectation values  with insertions of suitable local operators along the loop 
%represent 
 %correlators in this QFT and, 
at the conformal points
 obey the general properties of CFT correlators \cite{Drukker:2006xg}.
 For simple scalar operators that 
 are coupled to $\z$,  their insertions are 
  controlled by the dependence of 
 $\langle W^{(\z)}\rangle$ on $\z$.\footnote{
For recent examples of application 
of the defect CFT approach to Wilson loop computations in $\mc N=4$ SYM see 
\cite{Cooke:2017qgm,Giombi:2017cqn,Kim:2017sju,Giombi:2018qox}.}

%A non-perturbative RG flow connecting the two fixed points is expected at the non-%perturbative level.
%At weak-coupling it is captured by (\ref{1.2}). 

A  strong-coupling  counterpart  of the 
 RG flow  \rf{1.2}  was  discussed in 
\cite{Polchinski:2011im,Beccaria:2017rbe}. %This appears to be a quite interesting issue because
In particular,   AdS/CFT  predicts that  the locally BPS Wilson loop and the standard  Wilson loop 
should  be dual to the string partition  function  on the disc 
with   the 
standard (Dirichlet)  and  alternate  (Neumann)  boundary   conditions in $S^5$
\ci{Alday:2007he,Polchinski:2011im}. 
%of certain world-sheet fields. 
To  connect to the  strong-coupling limit  of the expectation   value 
of \rf{1.1} requires  finding  the weak-coupling series \rf{1.3}  to all orders in $\l$. 
Optimistically, this   might be possible  (despite the absence of  global supersymmetry for $\z\not=1$)  due to underlying integrability of the   $\mc N=4$ SYM   theory.

  \def \P  {\Phi}
  
   The  higher loop contributions to 
the  expectation value  of \rf{1.1}   simplify if one takes the 
large $\z$, small $\l $  limit with $\lambda\,\z^{2}\ll 1$.  In this case 
the scalar   coupling in \rf{1.1}  dominates over the vector one and also 
the planar SYM  theory becomes effectively  free  (with  only  the kinetic 
term for the scalar field $\P_m$ surving). As a result, the  planar 
scalar ladder diagrams  give  dominant   contribution to  $\langle W^{(\z)}\rangle$
 in this limit. Still, it is not clear  how    to find
 the exact (all-order) ladder-diagram expression for this expectation value 
 due to   non-trivial  effect of  the path ordering in \rf{1.1}.\footnote{The loop equation governing 
 the large $N$  adjoint scalar  loop expectation value  was  considered in \cite{Makeenko:1988hm}.
 }
    % if this  limit is exactly 
%Even in this 
%limit, the computation of $\langle W^{(\z)}\rangle$ is non-trivial and corresponds to a non-interacting
%scalar theory in large $N$ limit, hence with a planar.

One of our aims  below  will be to investigate how such ladder
 correlators are organized.
In \cite{Beccaria:2017rbe}, it  was  shown   that there is a rather simple 
 computational %and regularization 
procedure to  compute  ladder diagrams. We shall extend the 
discussion in  \cite{Beccaria:2017rbe}  to the $\l^3$  level. 
In general,  considering only ladder diagrams is  a major limitation, but  
one may attempt to reconstruct the full 
3-loop  $\l^3$  term in  (\ref{1.3}) by completing the ladder contribution  using  some  extra physical constraints.

We shall 
  also  explore   more general  loops with  non-constant 
$n^m$ in \rf{1.1}. 
%A related investigation concerns generalizations of the coupling 
%(\ref{1.1}).
In particular, one %simple possibility amounts to the introduction of a two
may consider a 2-parameter 
$\z=(\z_{1},\z_{2})$ family of   loop  operators 
$W^{(\z_{1},\z_{2})}$
 interpolating between the standard circular  Wilson loop and the 
$\frac{1}{4}$-BPS  WM loop considered in 
\cite{Zarembo:2002an,Drukker:2005cu,Drukker:2006ga}. 
This extension  is defined   by a circle  in $x^\m$-space and 
and $n^m$   corresponding to a latitude of $S^2  \subset S^5$:\foot{In what follows we shall often set the radius  of the circle $R$ to 1.}
% by 
%the following coupling to scalars
% (below  we set often set  the radius of the circle $R$ to 1)
\begin{align}\la{1.4}
&  x^\m = R\,(\cos\tau, \sin\tau, 0,0) \ ,  \qquad 
 n^m =(\sin \theta_0\cos \tau, \sin \theta_0\sin \tau, \cos \theta_0 , 0, 0, 0) \ , 
\\
&W^{(\z_1,\z_2)}:  \ \     \qquad 
\z  \Phi_{m} n^{m} = \z_{1}\,\Phi^{3}+\z_{2}\,(\cos\tau\,\Phi^{1}+\sin\tau\,\Phi^{2}) \ , \ \ \ \ \la{1.44} \\ 
&\qquad\qquad \qquad\qquad   \z_1 = \z  \cos \theta_0 , \ \qquad   \z_2 =\z \sin \theta_0 \ . \la{1.5}
\end{align}
For $\z =1$  this is the  $\frac{1}{4}$-BPS latitude WM loop for which the exact 
expression is   known  -- given by the    $\frac{1}{2}$-BPS   WM 
loop with the replacement $\l \to \l \cos^2 \theta_0$ \ci{Drukker:2006ga,Pestun:2007rz}.
  % where one has conformal invariance. 
  $W^{(\z_1,\z_2)}$   may be viewed as a 2-parameter deformation of the 
  $\frac{1}{4}$-supersymmetric (preserving     2  out of 8 $Q$-supercharges) 
    loop of  \cite{Zarembo:2002an} that has trivial   expectation value
    which corresponds to 
  the  special case  of    $\z_1=0,\z_2=1$  or $\z=1,\ \theta_{0}= {\pi \ov 2}$.
  Here we  shall analyze the renormalization of the couplings $\z_{1}$ and $\z_{2}$ at the two loops, and
at three loops in ladder limit.  This will provide interesting information about the two 
$\beta$-functions
associated to $\z_{1}$ and $\z_{2}$ generalizing (\ref{1.2}).

Another non-supersymmetric  Wilson loop generalization we shall study is 
%A third generalization we are going to discuss is the effect of 
the introduction of a 
non-trivial winding of the $n^m(\tau)$  contour   in the auxiliary  $S^{5}$ space.
% where $n^{m}$ lives. 
In particular, one  may  make  the replacement $\tau\to \nu\tau$ in  the expression for  $n^m$ in 
(\ref{1.4}). 
For $\z=1$ the corresponding WM loop  will no longer be $\frac{1}{4}$-BPS
and thus $\langle W \rangle$  will   be a non-trivial   function   %$\mc W(\nu,\l)$
of the winding  $\nu$ and  the  coupling $\l$ yet to be determined.\footnote{Note that  this deformation is different from multiply wound 
supersymmetric generalizations of the circular 
WM loop where  winding in space-time is correlated with that  in the 
scalar coupling term, see  \cite{Drukker:2007qr}.
}
% of $\nu$  and $\l$. 
%If this deformation is specialized to the globally 
%supersymmetric $\frac{1}{4}$-BPS loop, one may expect to obtain a non trivial function $\mc W(\nu)$
%of the winding (and of the coupling $\lambda$). 
%Since supersymmetry is still locally preserved, but 
%broken globally, the quantity $\mc W(\nu)$ may have nice special properties. 
We shall be compute it at the two-loop  order 
 demonstrating  its finiteness    (which is  expected to hold to all orders 
 as  this is a special  case of a locally supersymmetric WM loop).

\

%The plan of the paper is  following.
 We shall   start in  Sec.~\ref{sec:lad1} with a  review of an efficient regularization
procedure suitable for the analysis of the ladder diagrams
contributing to the expectation value of  generalized Wilson  loops. We  shall  apply
it at three loops  in the case of the circular contour 
% (\ref{1.1}) 
and discuss  renormalization properties of the resulting expectation value. 

In Sec.~\ref{sec:2par} we  shall  discuss the two-parameter loop (\ref{1.4}). We shall
present its complete two 
loop expression by adapting 
%the dimensional reduction calculations 
the results of \cite{Beccaria:2017rbe}
to this case.
 The three-loop contribution from ladder diagrams  will  also  be 
 found,  %discussed
 % and some information 
%is collected 
extracting   information about the $\beta$-functions 
for  the two couplings $\z_{1}, \z_{2}$.

Sec.~\ref{sec:wind} will be  devoted to the analysis of the effect of winding 
of the  scalar-space contour. 
We will   present  the two-loop expression of the wound loop by combining 
a direct computation of the ladder diagram 
contribution with an indirect determination
 of the non-ladder 
contribution  inferred by exploiting the 1d defect CFT perturbation theory. Several Appendices will 
contain  some technical details. % and related discussions. 

%\section*{\red{Comments about strong coupling}}
%
%\begin{enumerate}
%\item What can be said at strong coupling ? 
%Generalization of Zarembo's  solution to any $\nu$: this is
%straightforward  following 4.14  and below  in 0205160 --  we get
%induced metric
%$
%ds^2 =  ( 1 /\sinh^2 t   + \nu^2/\cosh^2  \nu t) ( dt^2  + ds^2)
%$
%the area is  still  same as in 4.20
%$= \int_{\eps}^{\infty }2 \pi ( \tanh \nu \tau - \coth  \tau) =
%2\pi/\eps + \mc O(\eps)$
%--   singular term   not depending on nu  without extra finite term
%depending on $\nu$. Thus    nu is not seen at leading order at  strong
%coupling -- renormalized area is
%still =0.
%
%\item
%connection  to sect 2.4.1 in 0711.3226:  -  our case  is  $k1=0$ in
%their solution
% while for full latitude there will be something else  as in their  eq 2.49
%
%\end{enumerate}
%

\def \btau {{\bm \tau}}
%%%%%%%%%%%%%%%%%%%%%%%%%%%%%%%%
\section{Ladder diagram contribution  to   $\z$-deformed  $1\ov2$-BPS circular WM loop}
\la{sec:lad1}

In this section we shall consider the expectation of the operator  (\ref{1.1})
for  the  standard circular   loop   with  $n^m=\const$ and discuss the evaluation of the planar ladder diagram
contributions. We shall use 
 a particular regularization  scheme based on  mode expansion and  convenient
point splitting.    
From a computational  point of view, this approach is superior to 
the dimensional regularization \cite{Beccaria:2017rbe}.
  This 
claim does not apply to non-ladder diagrams that will
 not be considered in this section.\footnote{In this case 
 the dimensional regularization (or, more precisely, 
regularization by dimensional reduction appropriate in  a supersymmetric theory)
is the most convenient  one.} 

\subsection{Mode regularization} % and point-splitting}

%We shall consider   planar ladder diagrams associated with a circular loop
%in space-time.
 Let us briefly recall the mode regularization method proposed in \cite{Beccaria:2017rbe}.
In the case of the $\z$-deformed  circular WM loop  contributions  from 
 $\ell$-loop planar ladder diagrams  containing  $\ell$   scalar  and  vector propagators  attached to the circular loop 
    lead to   expressions like 
\begin{align}
\la{2.1}
& I_{G_\ell}\equiv I_{(i_{1}i_{2})\dots (i_{2\ell-1} i_{2\ell})} = \int_{\tau_{1}>...>\tau_{2\ell}}
d^{2\ell}\bm\tau\  \mathcal G(\tau_{i_{1}i_{2}})... \mathcal G(\tau_{i_{2\ell-1}i_{2\ell}}), \\
&\mathcal G(\tau) = (\z^{2}-\cos\tau)\,\mathscr D(\tau), \qquad \mathscr D(\tau) \equiv  \frac{1}{4\,\sin^{2}\frac{\tau}{2}},\qquad \tau_{ij}=\tau_{i}-\tau_{j}.
\la{2.2} 
\end{align}
Here  $\{i_{1},\dots, i_{2\ell}\}$ is a permutation of $\{1, \dots, 2\ell\}$ associated with a 
%\underline
{planar} diagram $G_\ell =  ({i_{1}i_{2}})\dots  ({i_{2\ell-1}i_{2\ell}})$.
The diagram is built by taking points $\tau_{1}>\tau_2 >...>\tau_{2\ell}$ on the circular loop and 
connecting the pairs $(i_{1},i_{2})\dots (i_{2\ell-1},i_{2\ell})$. Clearly, the planarity constraint
allows only certain permutations.\footnote{Notice that there may be different pairings leading to the 
same diagram topology due to periodicity on the circle. An example is the two-loop 
equivalence $(12)(34)\simeq (14)(23)$. Nevertheless, summing over \underline{all} planar 
terms as in (\ref{2.1}) gives the correct contribution without possible over- or under-counting. This is 
because (\ref{2.1}) is nothing but enumeration of all possible contractions after the expansion of the exponential
in the loop operator.
}
In what follows   we shall use the following notation for  the path-ordered integral  in \rf{2.1}:
\be  \la{2.3}
\int [ d^{2\ell} \btau] \    F(\btau)   \equiv \int_{\tau_{1}>...>\tau_{2\ell}}
d^{2\ell}\btau\,  F(\btau)   = \int_0^{2\pi} d \tau_1 \int_0^{\tau_1} d \tau_2... \int_0^{\tau_{2\ell-1} } d \tau_{2\ell} \   F(\btau) \ . 
\ee
The mode regularization  procedure  is based on using   the    formal  Fourier mode
expansion  of\foot{$\mathscr D(\tau)$   is the scalar propagator (which is also the same as the  vector field propagator in the Feynman gauge that we shall assume) restricted to the  circle.}
  $\mathscr D(\tau)= {1 \ov 2 (1- \cos\tau ) }  =  -\sum^\infty _{n=1}   \, n\,\cos(n\,\tau) $     with a 
particular short-distance cutoff  $\eps \to 0$ \foot{Here $\eps$ is dimensionless  parameter,  i.e. 
$\eps = \eps' \mu$, where $\eps'\to 0$   is small-scale   cutoff 
and $\mu$ is a normalization  mass scale (e.g., the    inverse radius of the circle  
which was set to 1   in \ref{2.2}).}
%amounts to exploit the identity
\be \la{2.4}   % {1 \ov 4\,\sin^{2}  \frac{ \tau }{2}}    = -\sum^\infty _{n=1}    n\,\cos(n\,\tau) \ \ \  \to \ \ \ 
  {\mathscr D} (\tau) \ \to \ {\mathscr D}_\eps (\tau)   = {1-\cos\tau\, \cosh\eps \ov 2 (\cosh\eps - \cos \tau )^2  } =
   -\sum^\infty _{n=1} e^{-n \eps}  \, n\,\cos(n\,\tau)\ .
\ee 
Then we get the regularized expression for \rf{2.1}
\begin{align}
&I_{G_\ell}(\eps) = (-1)^{\ell}\,\sum_{n_{1}, \dots,  n_{\ell}=1}^{\infty}
e^{-\eps\,(n_{1}+...+n_{\ell})}\,
n_{1}\dots n_{\ell}\no  \\
&\times 
 \int[
d^{2\ell}\btau]\,  (\z^{2}-\cos\tau_{i_{1}i_{2}})\dots(\z^{2}-\cos\tau_{i_{2\ell-1}i_{2\ell}})\,
\cos(n_{1}\,\tau_{i_{1}i_{2}})\dots \cos(n_{\ell}\,\tau_{i_{2\ell-1}i_{2\ell}}).   \la{2.5}
\end{align}
Expanding in  $\eps\to 0$ and discarding poles and terms $\mc O(\eps)$, 
we are left with an expression that may contain powers of $\log\eps$. These  should be  the counterparts of  the 
dimensional regularization poles.  

Writing  first %For the actual calculation, it is convenient to exploit the identity 
\be\la{2.6} 
(\zeta^{2}-\cos\tau)\,\mathscr D(\tau) = (\zeta^{2}-1)\,\mathscr D(\tau)+\tfrac{1}{2},
\ee
 and then using \rf{2.4} 
 one may represent the contribution of   each diagram as  a power series in $\z^{2}-1$
\be
\la{2.7}
I_{G_\ell } = \sum_{r=0}^{\ell}(\zeta^{2}-1)^{r}\,I^{(r)}_{G_\ell} \ .
\ee
Here the $r=0$  term corresponds to the  $\frac{1}{2}$-BPS   WM  loop, i.e.     \ci{Erickson:2000af}   %independent on $G$ and reads 
\be
I^{(0)}_{G_\ell}  = \frac{1}{2^\ell}\ \frac{(2\pi)^{2\ell}}{(2\ell)!} =\te \{1, \pi^2,  \frac{\pi^{4}}{6}, \frac{\pi^{4}}{6}, 
\frac{\pi^{6}}{90},\dots \}, \ \ \ \ \ \ 
%\ \text{for}
\ \ \ \   \ell=0,1, 2,3,\dots. \la{2.8} 
\ee
Then 
\be  \la{2.9}
\langle W^{(\z)}\rangle = \sum_{\ell=0}^\infty \frac{\l^\ell }{(8\,\pi^{2})^{\ell}}  \sum_{G_\ell}   I_{G_\ell }  \ , \ee
where  the $\ell$-loop  % at  $\ell$ loops  is obtained  by adding a 
 normalization
prefactor
%$\frac{1}{2^{\ell}\,(4\,\pi^{2})^{\ell}}$ % = \frac{1}{(8\,\pi^{2})^{\ell}}$ 
 has  $1 \ov 2^{\ell}$ 
coming  from  the colour generator $t^{a}t^{a}$ contractions   and  $1\ov (4\pi^{2})^{\ell}$  from 
the normalization of the  scalar  field propagator.  

%%%%%%%%%%%%%%%%%%%%%%
\subsection{Three-loop contributions to  $\langle W^{(\z)}\rangle$}
%%%%%%%%%%%%%%%

The two-loop analysis can be found in \cite{Beccaria:2017rbe} and is briefly summarized in 
Appendix~\ref{sec:prelim}. We have only the two diagrams $(12)(34)$ and $(14)(23)$
and they give the contributions (dropping the  power-divergent $1\ov \eps^k$ terms)
\begin{align}
I_{(14)(23)} = \frac{\pi^{2}}{2}\,(\zeta^{2}-1)^{2}+\frac{\pi^{4}}{6} \ , \qquad \qquad
I_{(12)(34)} =2\pi^{2}\, (\zeta^{2}-1)\,\log\eps+\frac{\pi^{4}}{6}.\la{2.10} 
\end{align}
Including also the one-loop term $\frac{\lambda}{8}$ we obtain
\begin{align}
\langle W^{(\z)}\rangle_{\rm ladder} &= 1+\frac{\lambda}{8}
+ \lambda^{2}\,\Big[\frac{1}{192}+\frac{(1-\zeta^{2})^{2}}{128\,\pi^{2}}
{-\frac{1-\zeta^{2}}{32\,\pi^{2}}\log\eps}\Big] + \mc O(\l^3)\ .\la{2.11} 
\end{align}
The   logarithmically 
divergent term is canceled by  the contribution of other non-ladder diagrams  with internal   vertices; this  contribution   does not,  however, 
  change the  finite part (see  \cite{Beccaria:2017rbe}).  
 Consistently  with the fact that ladder   diagrams  should 
 dominate in the large $\z$ limit, 
   the divergent   term in \rf{2.11}   is subleading at large $\z$, {\em i.e.} it 
is $\mc O(\z^{2}\lambda^{2})$ as compared  to  the finite contribution $\mc O(\z^{4}\lambda^{2})$
that comes purely from the  ladder  diagrams. % without internal vertices.

\def \zetaR  {\zeta_{_{{\rm R}}}}

The novel three-loop contributions are derived in detail  in 
Appendix \ref{appA2},\ref{appA3}. The three-loop   ladder 
diagrams $G_3= (12)(34)(56)$, {\em etc.},  give the following contributions  $I^{(r)}_{G_3}$
to  \rf{2.1},(\ref{2.7})  (we omit $\mc O(\eps)$  terms)
\begin{align}
I^{(3)}_{(16)(25)(34)} &= 
\frac{1}{\eps}\Big(\pi^{2}\log\eps+\pi^{2}\log 2+\frac{\pi^{2}}{2}\Big)
-\frac{3\pi^{2}}{2}, %+\mc O(\eps)
\notag \\
I^{(2)}_{(16)(25)(34)} &= 
-\frac{\pi ^4}{6\, \eps }+\frac{\pi ^4}{2},%+\mc O(\eps)
\notag \\
I^{(1)}_{(16)(25)(34)} &= 0,
\notag \\
  %\text{\rule[0.45ex]{0.5\textwidth}{0.5pt}} \notag \\
I^{(3)}_{(12)(34)(56)} &= 
\frac{\pi^{2}}{2} \, \log \eps +\frac{\pi^{2}}{2} \, \log 3,%+\mc O(\eps)
\notag \\
I^{(2)}_{(12)(34)(56)} &= 
3 \pi ^2 \log ^2\eps, % +\mc O(\eps)
\notag \\
I^{(1)}_{(12)(34)(56)} &=
\frac{1}{2} \pi ^4 \log \eps + \frac{3 \pi ^2}{2} \zetaR (3),%+ \mc O(\eps)
\notag \\
 % \text{\rule[0.45ex]{0.5\textwidth}{0.5pt}} \notag \\
I^{(3)}_{(12)(36)(45)} &=  I^{(3)}_{(14)(23)(56)} = 
-\frac{\pi^{2}\log\eps}{2\,\eps}+\frac{\pi^{2}}{2}\log\eps+\frac{\pi^{2}}{4}, %+\mc O(\eps)
\notag \\
I^{(2)}_{(12)(36)(45)} &= I^{(2)}_{(14)(23)(56)} = 
-\frac{\pi ^4}{6 \eps }+\pi ^2 \log ^2\eps -\pi ^2 \log \eps+\frac{\pi ^2}{2},
%+\mc O(\eps)
\notag \\
I^{(1)}_{(12)(36)(45)} &=  I^{(1)}_{(14)(23)(56)} = 
\frac{1}{3} \pi ^4 \log  \eps    + 2 \pi ^2 \zetaR (3),
%+\mc O(\eps)
\notag \\
 % \text{\rule[0.45ex]{0.5\textwidth}{0.5pt}} \notag \\
I^{(3)}_{(16)(23)(45)} &= 
\frac{4\pi^{2}}{3\eps}\log 2 +\frac{\pi^{2}}{2}\log \eps
-\frac{\pi^{2}}{2}\log 3-\frac{2\pi^{2}}{3},%+\mc O(\eps)
\notag \\
I^{(2)}_{(16)(23)(45)} &= 
-\pi ^2 \log ^2\eps  +   \frac{\pi ^4}{6},%+\mc O(\eps)
\notag \\
I^{(1)}_{(16)(23)(45)} &= 
\frac{\pi ^4}{6}  \log \eps -\frac{\pi ^2 }{2}\zetaR (3).
%+\mc O(\eps)
\la{2.12}
\end{align}
Here  $\zetaR(n)$   is the Riemann   zeta-function. 
The  three-loop  
ladder contributions to the Wilson loop \rf{1.1},\rf{2.11}   are then    (dropping power divergences) 
\begin{align}
\la{2.13}
\langle W^{(\z)}\rangle_{\rm ladder} = 1&+\frac{\lambda}{8}
+ \lambda^{2}\,\Big[\frac{1}{192}+\frac{(1-\zeta^{2})^{2}}{128\,\pi^{2}}
{-\frac{1-\zeta^{2}}{32\,\pi^{2}}\log\eps}\Big]\notag  \\
&+\lambda^{3}\,\Big[
\frac{1}{9216}-\frac{5\,(1-\zeta^{2})}{512\,\pi^{4}}\,\zetaR(3)  +\frac{(1-\zeta^{2})^{2}}{768\,\pi^{2}} \Big( 1 
+ \frac{1}{2\,\pi^{2}} (8-5\,\zeta^{2}) \Big) \notag \\
&{- \Big( \frac{1-\zeta^{2}}{384\pi^{2}}
%\,\log\eps
 + \frac{(1-\zeta^{2})^{2}\,(2-\zeta^{2})}{256\,\pi^{4}}\Big) \log\eps+\frac{(1-\zeta^{2})^{2}}{128\,\pi^{4}}\log^{2}\eps}\Big]  + \mc O(\l^4)\ .
\end{align}

\subsection{Discussion}

If we pretend  for a moment 
that the ladder approximation is consistent by itself, we find that 
the logarithmically  
divergent terms   in (\ref{2.13})
may be  {\it formally} absorbed  into a redefinition   of both $\z$ and $\l$,\foot{Some motivation for doing this  may be as follows: 
if we consider just a free   theory of adjoint vectors and scalars   and then compute  the expectation of the generalized WL
\rf{1.1}  one  may expect   renormalizability  -- then one may be allowed  to formally renormalize both $\z$ and $\l$.} 
%But this does not seem to have a deeper meaning.}
{\em i.e.} assuming that  the parameters appearing in \rf{2.13}  are the   "bare"   ones  
%({\bf   may be omit this or make this consistent with  right 1-loop $\beta$-function ??}) 
%by taking into account that $\lambda, \zeta$ are {\em bare} ones in the  above, and writing
\begin{align}
\la{2.14} \zeta_{\rm b}(\eps)  &= \zeta+\frac{\lambda}{8\,\pi^{2}}\,\zeta\,(\zeta^{2}-1)\,\log\eps+\dots \ , \\
\lambda_{\rm b}(\eps) 
 &= \lambda+\frac{\lambda^{2}}{4\,\pi^{2}}\,(1-\zeta^{2})\,\log\eps
+\frac{\lambda^{3}}{32\,\pi^{4}}\,\Big[2\,(1-\zeta^{2})\,\log^{2}\eps
+(1-\zeta^{2})^{3}\,\log\eps\Big]+\dots \ .  \la{2.15}
\end{align}
Here  one  may  make the dependence of the renormalized parameters $\z, \l$ 
 on the renormalization  mass scale  $\mu$ explicit  by redefinining 
 $\eps \to \eps \mu$ % ($\muMR$   stands for  mode regularization scheme) 
and setting $ {d\lambda_{\rm b} \ov d \mu} =0, \  
{d\z_{\rm b} \ov d \mu} =0$. 
Note that the sign of $d\z\ov d \mu$  
here  appears to  be  opposite to the one in the $\beta$-function \rf{1.2}.

The  t' Hooft coupling $\l$  should not of course run in the {\em full}  SYM  theory  %   but here we consider just a subclass of ladder graphs.}  
(which includes interactions and thus  also diagrams  with internal vertices) 
so  it should not run in (\ref{2.13}).  In fact,  all logarithms 
  should be  cancelled   by  ({\em i})  the  expected   1-loop 
renormalization of $\z$    consistent with \rf{1.2}, and ({\em ii}) all    remaining divergences
 should be cancelled by the contributions of other diagrams. 
%we should then  have for the $\z$-renormalized 
%This (minimal) renormalization 
%removes all red terms in $\langle W\rangle_{\rm ladder}$.
\iffa 
 \footnote{  ????? Not sure what is the meaning   of this expression --  2-loop   log should then be kept ? -- it is cancelled by other diagrams....
If all divs  are cancelled by other diagrams:
\begin{align}
\langle W^{(\z)}\rangle_{\rm ladder} &= 1+\frac{\lambda}{8}
+ \lambda^{2}\,\Big[\frac{1}{192}+\frac{(1-\zeta^{2})^{2}}{128\,\pi^{2}}
\Big]\notag \\
&+\lambda^{3}\,\Big[
\frac{1}{9216}-\frac{5\,(1-\zeta^{2})}{512\,\pi^{4}}\,\zetaR(3) + \frac{5 (1-\zeta^{2})^{2}}{1536\,\pi^{2}}
+ \frac{5  (1-\zeta^{2})^{3}\,}{1536\,\pi^{4}}
\red{+\frac{(1-\zeta^{2})^{2}\,\zeta^{2}}{256\,\pi^{4}}\,\log\eps} \Big] \no 
\end{align}
}
\fi 
%\medskip
%\noindent
A major simplification occurs if we 
decide to  keep only  the highest power of $\z$ at each order in expansion in $\l$.  
These  leading terms  may get contributions   only from the  {\it scalar}  ladders graphs
(ladder graphs   with vector propagators give  terms  subleading in $\z$
which also receive contributions from other non-ladder diagrams). Thus,
 it  should   be captured exactly  by  the expression in \rf{2.13}, {\em i.e.} 
 \be
\la{2.16}
\langle W^{(\z)} \rangle_{\rm \zeta\gg 1} = 1 + {\l \ov 8}   +   
 \frac{\,\lambda^{2}\zeta^{4}}{128\,\pi^{2}}+
\frac{\lambda^{3}\, \zeta^{6}}{1536\,\pi^{4}} (-5{+6\,\log\eps}) + \OO(\l^4) \   . 
\ee
The coupling in (\ref{2.16}) is the {\em bare} one $\z_{\rm b}$ and the divergence 
is canceled by the redefinition in terms of the renormalized $\z_{\rm r}$
consistent with the  beta function
%v2 
in (\ref{1.2})  (cf. \rf{2.14})\footnote{To simplify the  expressions, 
in the following we shall  usually suppress  the ``b'' (bare) and ``r'' (renormalized) 
subscripts  on $\z$ in respective  expectation  values. 
%couplings
%and will not write them explicitly.
 %No ambiguities should arise since, generally speaking, 
In general, the bare couplings will always be accompanied by $\log\eps$ terms, while the  renormalized couplings
will come together with $\log\mu$ terms.} 
% (in contrast to \rf{2.14})
%v2
\be \la{ren}
\zeta_{\rm b} = \zeta_{\rm r} -\frac{\lambda \zeta^{3}_{\rm r}}{8\,\pi^{2}}\,\log (\mu\, \eps) +
\mc O(\lambda^{2}).
\ee
% involving the renormalization scale.}
This is of course not unexpected as  the $\z^3$ term in the $\beta$-function \rf{1.2}   comes 
from the ladder  graph with the  scalar   propagator   \cite{Polchinski:2011im}.
After the renormalization, the divergent $\log\eps$ factor  is  replaced by $\log\mu$, {\em i.e.}
the renormalized expression reads
% where $\mu$ is the
%renormalization scale. The renormalized loop is then
\be
\la{2.17}
\langle W^{(\z)} \rangle_{\rm \zeta\gg 1} = 1 + {\l \ov 8}   +   
 \frac{\,\lambda^{2}\zeta^{4}}{128\,\pi^{2}}+
 %v2
\frac{\lambda^{3}\, \zeta^{6}}{1536\,\pi^{4}} (-5{-6\,\log\mu}) + \OO(\l^4) \   . 
\ee
%The sign of the log term  here is consistent with beta function in (\ref{1.2})  (cf. \rf{2.14}). 
 %This is because in the large $\zeta$ limit non-ladder diagrams are absent, $\lambda$ is not renormalized and we are working in the 
%correct theory.  Instead, at subleading order in $\zeta$, the missing non-ladder diagrams are responsible
%for the running of $\lambda$ and this induces an extra running in $\zeta$ changing the sign of the log terms
%when comparing (\ref{2.14}) and (\ref{4.6}).
%\medskip \noindent
%Just to examine the structure, 

If  one  includes also the  4-loop  ladder  graph contributions and   keeps 
 only the leading $\l^n \z^{2n}$  terms
one expects to find  %the following structure 
 \begin{align}
\langle W^{(\z)} \rangle_{\rm \zeta\gg 1}
 =  
1  &+ {\l\ov 8} +  \frac{\lambda^{2}\zeta^{4}}{128\,\pi^{2}} +
\frac{\lambda^{3}\zeta^{6}}{1536\,\pi^{4}} (-5{+6\,\log\eps}) \no \\   &+ 
\frac{\l^4 \zeta^{8}}{4096\,\pi^{6}} \big[    w_4 {+2\,(-5 + b_{2})\,\log\eps+9\,\log^{2}\eps }
\big]\,+\OO(\l^5) \ . \la{2.19} 
\end{align}
Here $w_4$ is a finite   constant  and the $\log^{2}\eps$  term   
is fixed by   consistency with the  1-loop 
$\beta$-function \rf{1.2}, {\em i.e.}  it should be possible to eliminate 
all divergences  by the  following    redefinition $\z \to \z_{\rm b} (\eps,\z)$
in \rf{2.19} (here $\z=\z(\mu)$ is the renormalized coupling) 
\be\la{2.20}
%v2
\zeta_{\rm b} = \zeta -\big[\frac{\l\zeta^{3}}{8\pi^{2}}+ b_{2}\frac{\l^2\zeta^{5}}
{(8\pi^{2})^{2}}+\dots\big]\,\log (\mu\, \eps)  +     { 3 \l \z^5 \ov 128 \pi^4}\, \log^2  (\mu\, \eps) + ...\ , 
\ee
where $b_{2}$ is  the 2-loop   coefficient in the $\beta$-function. In the  large $\z$ limit
\be \la{2.21}
%v2
\beta(\z\gg 1 ) =  \frac{\lambda \zeta^{3}}{8\,\pi^{2}} +   \frac{\lambda^2 \zeta^{5}}{(8\,\pi^{2})^2} b_2 +  \OO(\l^3) \ . \ee
%scheme dependent, {\em i.e.} it is shifted if $\eps$ is rescaled by some numerical factor. 
Direct computation  of the 4-loop term in \rf{2.19}  remains a challenge.
In the next section we will  attempt to indirectly infer additional information about the value of $b_2$ 
by considering a more general Wilson loop.

%%%%%%%%%%%%%%%%%%%%%%%%%%%%%%%%%%%%%%
\section{$\z$-deformation  of  $\frac{1}{4}$-BPS  { latitude}  WM loop}
\la{sec:2par}
%%%%%%%%%%%%%%%%%%%%%%

Let us   now  consider a generalization of  the   $\frac{1}{4}$-BPS 
supersymmetric  Wilson-Maldacena  loop  corresponding to latitude in $S^5$ \cite{Zarembo:2002an,
Drukker:2005cu,Drukker:2006ga}  defined by \rf{1.4}
where $\theta_0$ is  a constant parameter.
% Its  space-time  part is   defined  again by a circle 
%$(x^{1},x^{2}) = (\cos\tau, \sin\tau)$   but  the scalar coupling vector $n^m$   in \rf{1.1}  is no longer  constant:
%but it couples to the three scalars $\Phi_{1}$, $\Phi_{2}$, and $\Phi_{3}$ according to 
%the usual term $|\dot x(\tau)|\,n^{I}(\tau)\,\Phi_{I}$ with the following
%latitude angle dependent coupling
\iffa \be
n^{1}=\sin\theta_{0}\,\cos\tau,\qquad 
n^{2}=\sin\theta_{0}\,\sin\tau,\qquad 
n^{3}=\cos\theta_{0}. \la{3.1} 
\ee
\fi 
%The  WM  loop  corresponding to \rf{1.4}   has a special 
%feature of  being  $\frac{1}{4}$-BPS. % with respect to global supersymmetry.
Due to  $\frac{1}{4}$-BPS  property of this WM loop, the   dependence 
of its expectation  value on the latitude angle $\theta_0$ 
can be  found just by   the  redefinition  $\lambda\to \lambda\,\cos^{2}\theta_{0}$  in the expectation 
value for the  $\ha$-BPS  circular   WM  loop   \cite{Drukker:2006ga,Pestun:2009nn}. 
As in the circle case   \ci{Erickson:2000af}
  all  non-vanishing contributions  to the latitude WM loop come from ladder diagrams while  contributions of 
non-ladder diagrams mutually cancel. 
For $\theta_{0}=0$ we get  
back the $\frac{1}{2}$-BPS circle where $\langle W\rangle = { 2 \ov \sqrt \l}  I_1 ( \sqrt \l)$ while for 
 $\theta_{0}= {\pi\ov 2}$  we get  the $\frac{1}{4}$-supersymmetric
 %$\frac{1}{4}$-BPS 
 loop of  \cite{Zarembo:2002an}   for which 
$\langle W\rangle=1$.
%\foot{This is a consequence of  the residual   global 
%supersymmetry and can be seen explicitly 
%as a consequence of the  vanishing  of the effective zero coupling  $\lambda\,\cos^{2}\theta_{0} \to 0$.}

To   generalize 
%Similar to the $\frac{1}{2}$-BPS loop, we again generalize
 the    latitude  loop we  add a coefficient $\z$
in front of the scalar coupling term  as in \rf{1.1},\rf{1.44}. This is  also equivalent 
 to introducing  the two couplings $\z_1,\z_2$  as in \rf{1.5}.
\iffa
\be\la{3.2} 
\zeta_{1} =\z\,  \cos \theta_0\ ,\qquad \zeta_{2} =\z\,  \sin \theta_0 \ .   %\,\zeta,\qquad (\text{c}=\sqrt{1-\text{s}^{2}}).
\ee
\fi
Below  we shall  first present the full two-loop expression for  $\langle W^{(\z_{1},\z_{2})}\rangle$ defined  by   \rf{1.1}  and  (\ref{1.44})  and 
then  discuss the three-loop contributions from ladder diagrams only.
%in the $\z\gg 1$ regime, {\em i.e.}
%keeping terms of the form $\lambda^{\ell}\,\z_{1}^{n}\z_{2}^{2\ell-n}$ with $n=0, \dots, 2\ell$.

Let us note that  for $\tau$-dependent direction $n^{m}$, 
 there is no 1d reparametrization  and, in particular, scale  invariance  in the 
 WM loop \rf{1.1}. Thus for  $\theta_0\not=0$ there 
 will    no 1d conformal invariance  even  for $\z=0$ or $\z=1$.\foot{In particular, 
%As a technical comment, we remark is that 
%for generic values of $\theta_{0}$, the latitude loop is not scale invariant because
%the coupling $\int dt\,[A_{\mu}\dot x^{\mu}+|\dot x|\,\Phi_{I}\,n^{I}(\tau)]$
%is invariant under reparametrizations of $\tau$ only if $n^{I}$ is constant.
computing correlation functions  of scalar fields along the loop, i.e.
 $\langle\text{Tr}[\Phi\cdots\Phi\,\exp\int(i\,A\cdot \dot x+|\dot x|\Phi\cdot n)]\rangle$,
we cannot interpret them as 1d CFT correlators because of the explicit $\tau$ dependence in the 
scalar coupling in the exponent.}
The explicit classical  breaking of  scale  invariance %is not a priori incompatible
is   not in conflict with UV finiteness that  still holds due to 
local  supersymmetry of the WM loop  ($\z=1$) case. 
%Indeed, we know that for $\z=1$ and generic $\theta_{0}$, the WM latitude loop  is 
%UV finite and thus $\beta_{i}(\z_{1},\z_{2})=0$. 
Conformal perturbation theory will of course apply if we expand near 
the $z=0$    point, {\em i.e.} in powers of $\z_1$ and $\z_2$.

%Of course, nothing prevents from expanding 
%in $\z_{1}, \z_{2}$ around special conformal points where the defect CFT machinery %may be perturbatively applied.

\subsection{Complete two-loop contribution}

At two-loop level, it is possible to compute $\langle W^{(\z_{1},\z_{2})}\rangle$ by building 
on the analysis of the $\z$-deformation of the  circular  loop  in \cite{Beccaria:2017rbe}. 
In this case  the one and two-loop diagrams contributing  to $\langle W^{(\z)}\rangle$
% we see that the $\z$-coupling 
  contain the $\z$-coupling   in  the integrand factors like  % expression
\be
\la{3.3}
\z^{2}\,|\dot x(\tau_{1})|\,|\dot x(\tau_{2})|-\dot x(\tau_{1})\cdot \dot x(\tau_{2}) = 
\z^{2}-\cos\tau_{12}\ ,
\ee
where the first term is from the coupling of the scalars to the loop,
while the second term  corresponds to the  vector  coupling.  There is  one such factor in the one loop diagram
and in the two-loop self-energy and internal vertex
diagrams, and two such factors in two-loop ladder diagrams. 

To see this in  detail, let us adopt the same 
labeling of diagrams as in \cite{Beccaria:2017rbe} and  review  each 
contribution in the circular case separately. In dimensional regularization  with space-time dimension 
$d=2\,\omega \equiv 4-2\,\eps$, the only one-loop diagram is
\be
W_{1}(\z) = \frac{\Gamma(\omega-1)}{16\,\pi^{\omega}}\oint_{C} d^{2}\bm{\tau}\ 
\frac{\z^{2}\,|\dot x(\tau_{1})|\,|\dot x(\tau_{2})|-\dot x(\tau_{1})\cdot \dot x(\tau_{2})}{
|x(\tau_{1})-x(\tau_{2})|^{2\,\omega-2}}  \,
\ee
which  indeed contains the explicit factor (\ref{3.3}). 
%%%%%%%%%%%
At two loops, we have ladder, self-energy,  and internal
vertex contributions. The ladder diagrams are 
\begin{align}
W_{2,1a}(\z) &= \frac{\big[\Gamma(\omega-1)\big]^{2}}{64\,\pi^{2\omega}}\,\oint
%_{\tau_{1}>\tau_{2}>\tau_{3}>\tau_{4}}
[d^{4}\bm\tau]\,\frac{(\z^{2}\,|\dot x^{(1)}|\,|\dot x^{(2)}|-\dot x^{(1)}\cdot \dot x^{(2)})
(\z^{2}\,|\dot x^{(3)}|\,|\dot x^{(4)}|-\dot x^{(3)}\cdot \dot x^{(4)})}
{(|x^{(1)}-x^{(2)}|^{2}\,|x^{(3)}-x^{(4)}|^{2})^{\omega-1}}, \notag \\
W_{2,1b}(\z) &= \frac{\big[\Gamma(\omega-1)\big]^{2}}{64\,\pi^{2\omega}}
\,\oint%_{\tau_{1}>\tau_{2}>\tau_{3}>\tau_{4}}
[d^{4}\bm\tau]\,\frac{(\z^{2}\,|\dot x^{(1)}|\,|\dot x^{(4)}|-\dot x^{(1)}\cdot \dot x^{(4)})
(\z^{2}\,|\dot x^{(2)}|\,|\dot x^{(3)}|-\dot x^{(2)}\cdot \dot x^{(3)})}
{(|x^{(1)}-x^{(4)}|^{2}\,|x^{(2)}-x^{(3)}|^{2})^{\omega-1}}, \la{33}
\end{align}
{\em i.e.}  both have two factors of  (\ref{3.3}). The self-energy contribution   has one factor of \rf{3.3}
\be
W_{2,2}(\z) = - \, \frac{[\Gamma(\omega-1)]^2}{128\,\pi^{2\,\omega}(2-\omega)(2\omega-3)}\, 
\oint d\tau_{1}d\tau_{2}\,
\frac{\z^{2}|\dot x(\tau_{1})|\,|\dot x(\tau_{2})|-\dot x(\tau_{1})\cdot \dot x(\tau_{2})}{
\big[|x(\tau_{1})-x(\tau_{2})|^{2}\big]^{2\omega-3}}  \, .\la{34}
\ee
%with one factor (\ref{3.3}).
 Finally,  the sum of the internal vertex diagrams, one with
mixed  scalar-vector $\Phi\Phi A$ vertex and the other with a triple vector $A^{3}$ vertex, reads
(here $\Delta(x) = (-\del^2)^{-1} = \frac{\Gamma(\omega-1)}{4\pi^{\omega}} { 1 \ov |x|^{2 \omega -2}}$)
\begin{align}
W_{2,3}(\z) &= -\frac{1}{4}\,\oint d^{3}\bm{\tau}  \, \varepsilon(\tau_{1},
\tau_{2},\tau_{3})\ \Big[\z^{2}\,|\dot x^{(1)}|\,|\dot x^{(3)}|-\dot x^{(1)}\cdot \dot x^{(3)}\Big]\notag \\
&\ \qquad \qquad \times \dot x^{(2)}\cdot \frac{\partial}{\partial x^{(1)}}\int d^{2\omega}y\,
\Delta(x^{(1)}-y)\,\Delta(x^{(2)}-y)\,\Delta(x^{(3)}-y),\la{35}
\end{align}
and thus also    has   one factor  of (\ref{3.3}). 

Turning now to  the case of the latitude loop (\ref{1.44}), the combination (\ref{3.3})
is to be replaced by
\be
\la{3.8}
\z_{1}^{2}+\z_{2}^{2}\cos\tau_{12}-\cos\tau_{12} = (1-\z_{2}^{2})\,
\big(\widehat\z^{2}-\cos\tau_{12}\big),\qquad \widehat \z = \frac{\z_{1}}{\sqrt{1-\z_{2}^{2}}}.
\ee
Thus, we obtain 
\begin{align}
\la{3.9}
\langle W^{(\z_{1}, \z_{2})}\rangle &= 1+\lambda\,(1-\z_{2}^{2})\,
W_{1}(\widehat\z)\notag \\
&+\lambda^{2}\,\Big[
(1-\z_{2}^{2})^{2}\,W_{2,1}(\widehat\z)+
(1-\z_{2}^{2})\,W_{2,2}(\widehat\z)+
(1-\z_{2}^{2})\,W_{2,3}(\widehat\z)
\Big]+\dots.
\end{align}
Here the $W$-functions  are the ones  for the circle  case \rf{33},\rf{34},\rf{35}
 %contributions to (\ref{3.9}) have been
  computed  already in  \cite{Beccaria:2017rbe}.
Introducing the shortcut \be 
{1\ov \eps' }\equiv  {1\ov \eps}+2\log\pi+2\,\gamma_{\text{E}} \ , \la{36}\ee
they read\footnote{Notice that the sum of the non-ladder two-loop diagrams is proportional to $1-\z^{2}$ as it should be since
for $\z=1$ (i.e. the  
$\frac{1}{2}$-BPS loop)  it is known  that non-ladder diagrams mutually cancel at all orders.
}
\begin{align}
W_{1}(\z) &= \frac{1}{8}+\frac{1}{8}\,(1-\z^{2})\,\eps+\mc O(\eps^{2}), \notag \\
W_{2,1}(\z) &= \frac{1}{192}+(1-\z^{2})\,\Big[\frac{1}{64\,\pi^{2}\,\eps'}
+\frac{1}{128\,\pi^{2}}\,(7-3\,\z^{2})\Big]+\mc O(\eps),
\notag \\
W_{2,2}(\z) &= \z^{2}\,W_{2,2}{(1)}+(1-\z^{2})\,\Big[-\frac{1}{64\,\pi^{2}\,\eps'}
-\frac{1}{16\,\pi^{2}}\Big]+\mc O(\eps),\notag\\
W_{2,3}(\z) &= -W_{2,2}{(1)}+(1-\z^{2})\,\Big[-\frac{1}{64\,\pi^{2}\,\eps'}-\frac{1}{64\,\pi^{2}}
\Big]+\mc O(\eps),\notag\\
W_{2,2}{(1)} &=  -\frac{1}{64\,\pi^{2}\,\eps'}-\frac{1}{32\,\pi^{2}}+\mc O(\eps)  \ . 
\end{align}
Using these expressions  in (\ref{3.9}), we obtain 
\begin{align}
\la{3.11}
& \langle W^{(\z_{1}, \z_{2})}\rangle = 1+\frac{\lambda}{8}\,\Big[
1-\z_{2}^{2}+\eps\,(1-\z_{1}^{2}-\z_{2}^{2})+\mc O(\eps^{2})\Big]\notag \\
& +\lambda^{2}\Big[
\frac{1}{192}(1-\z_{2}^{2})^{2}+\frac{(\z_{1}^{2}+\z_{2}^{2}-1)(3\z_{1}^{2}+7\z_{2}^{2}-1)}
{128\,\pi^{2}}{ +\frac{\z_{2}^{2}\,(\z_{1}^{2}+\z_{2}^{2}-1)}{64\,\pi^{2}\,\eps'}}
+\mc O(\eps)
\Big]+\mc O(\lambda^{3}).
\end{align}
The couplings in  (\ref{3.11})   are the bare ones, {\em i.e.}
   they    should   be replaced by\footnote{The coupling $\lambda$ does not renormalize, but, as  discussed in  \cite{Beccaria:2017rbe},  the explicit factor $\mu^{2\eps}$ is 
nevertheless necessary to fix dimensions at generic $\eps$.}, cf. (\ref{3.13}),
\begin{align}
\la{3.12}
\lambda_{\text{b}} &= \mu^{2\,\eps}\,\lambda,\no   \\
\z_{1\,\text{b}} &= \z_{1}+\frac{1}{2\,\eps}\,\beta_{1}(\z_{1},\z_{2})
+\dots, \qquad
\z_{2\,\text{b}} = \z_{2}+\frac{1}{2\,\eps}\,\beta_{2}(\z_{1},\z_{2})
+\dots,\end{align}
where  $\z_i$  are the renormalized couplings. The resulting expression for \rf{3.11}   
expressed in terms of renormalized couplings   should   be finite    and that   determines 
%in    and  $\beta_i$. are the corresponding  RG   $\beta$-functions.
  %determined by requiring finiteness. 
%  This predicts t
  the leading contribution to the 
$\beta_{2}$-function  to be $\frac{\lambda}{8\pi^{2}}\,\z_{2}\,(\z_{1}^{2}+\z_{2}^{2}-1)$.
 We shall assume  that  $\beta_{1}$  has a similar   structure, {\em i.e.} 
\begin{align}\la{3.13}
\beta_{1}(\z_{1},\z_{2}) & %\stackrel{?}
{=} \frac{\lambda}{8\pi^{2}}\,\z_{1}\,(\z_{1}^{2}+\z_{2}^{2}-1)+\dots,  \qquad    %\notag\\
\beta_{2}(\z_{1},\z_{2}) & %\stackrel{?}
{=} 
\frac{\lambda}{8\pi^{2}}\,\z_{2}\,(\z_{1}^{2}+\z_{2}^{2}-1)+\dots.
\end{align}
This is a natural  generalization  of the  expression  for the  $\beta$-function for $\z$ in  \ci{Polchinski:2011im} found  for the 
$\z_1=\z, \, \zeta_2=0$ case. 
Indeed, assuming that 
the  diagrams  with internal vertices do not contribute to the  1-loop beta function 
we then need to add just the  ladder graph   with the vector  propagator   and that  leads to the  -1  terms in  \rf{3.13}.
% the above $\beta$-functions. 
Then  the   $\beta$-function for $\zeta=\sqrt{ \zeta_1^2 + \zeta_2^2}$ is  the same as in \rf{1.2} 
while  $\z_{1}/\z_{2}$ or $\theta_0$   in \rf{1.5} is not renormalized at one-loop order.

From (\ref{3.12}) and (\ref{3.13})   we obtain the finite expression ($\z^{2}=\z_{1}^{2}+\z_{2}^{2}$, see \rf{1.5})
%\begin{align}
%\la{3.14}
%\langle W^{(\z_{1}, \z_{2})}\rangle &= 1+\frac{\lambda}{8}\,(
%1-\z_{2}^{2}) +\lambda^{2}\Big[
%\frac{1}{192}(1-\z_{2}^{2})^{2}+\frac{(\z_{1}^{2}+\z_{2}^{2}-1)(\z_{1}^{2}+5\z_{2}^{2}-1)}
%{128\,\pi^{2}}\notag \\
%&+\z_{2}^{2}\,(\z_{1}^{2}+\z_{2}^{2}-1)\frac{\log(\overline\mu\,R)}{32\,\pi^{2}}
%\Big]+\mc O(\lambda^{3}).
%\end{align}
\begin{align}
\la{3.14}
\langle W^{(\z_{1}, \z_{2})}\rangle &= 1+\frac{\lambda}{8}\,(
1-\z_{2}^{2}) +\lambda^{2}\Big[
\frac{1}{192}(1-\z_{2}^{2})^{2}+\frac{(\z_{1}^{2}+\z_{2}^{2}-1)(\z_{1}^{2}+a\,\z_{2}^{2}-1)}
{128\,\pi^{2}}\Big]+\mc O(\lambda^{3}), \notag \\
a &\equiv  5+4\,\log(\muDR\,R),\qquad\qquad \muDR \equiv  \pi\,e^{\gamma_{\text{E}}}\, \mu\,,
\end{align}
where the presence of the  scheme-dependent constant $a$ reflects
 the running of $\z_{2}$ (DR refers to the dimensional regularization scheme).\footnote{The scheme where $a=1$ has the  special property that 
$\partial_{\z_{1}} \langle W^{(\z_{1},\z_{2})}\rangle$  gives the  $\beta$-function
for the coupling $\z_{1}$. In general, this is what happens for a perturbation of the free energy
$F$ around a conformal fixed point, {\em i.e.} we expect to have relations
$\partial_{g_{i}}F(\bm{g}) = C_{ij}(\bm{g})\,\beta_{j}(\bm{g})$
expressing stationarity of the free energy. Here, such a relation appears to be accidental
as  $\langle W^{(\z_{1},\z_{2})}\rangle = e^{-F}$ 
does not,  in  general,  have such an intepretation unless we are on 
the critical line $\z=1$.
}
Notice also that 
we have reintroduced the implicit length scale $R$.\footnote{
$x(\tau)\sim R$ implies that each loop comes with 
 a factor $\lambda_{\rm b} R^{2\,\eps}$ from $(|\dot x||\dot x'|-\dot x\cdot\dot x')/|x-x'|^{2\omega-2}$.
In terms of the renormalized coupling this is $\lambda\,(\mu\,R)^{2\,\eps}$ producing logs of the 
adimensional quantity $\mu\,R$.
}
%and redefined
%the renormalization scale $\overline\mu = \pi\,\mu\,e^{\gamma_{\text{E}}}$. 
Of course, for $\z=1$, we recover the expected  modification 
 $\lambda\to \lambda (1-\z_{2}^{2})=\l \cos^2\theta_0 $
of the $\frac{1}{2}$-BPS loop. The small $\theta_{0}$ expansion of (\ref{3.14}) is briefly discussed
in Appendix~\ref{sec:theta}.

\subsection{Three-loop ladder diagram contribution and the large $\z$ limit }

Restricting  to {ladder diagram} contributions, the expectation value  of 
the  loop   with  % $\frac{1}{4}$-BPS loop with 
generic couplings $\z_{1}$ and $\z_{2}$   can be  effectively 
obtained from the  knowledge of   expectation   value of $\z$-deformation of the 
$\frac{1}{2}$-BPS circular loop. According to  the previous discussion (cf. (\ref{3.8}))
to get the  expectation value   $\langle W^{(\z_1,\z_2)}\rangle_{\rm ladder}$ 
in the latitude  case  it is enough to 
make the following replacements  in (\ref{2.13})
%
% 
%This key observation   follows    the fact that  due to the special   "circle-like" 
% $\tau$-dependence  of the scalar coupling in \rf{1.4},\rf{3.1} 
%the numerators in generic ladder diagrams 
%with  $\ell$ propagators are  simply related to those in the circular   loop case 
%  (denominators do not depend on $\z$, cf. \rf{2.1}): 
%\begin{align}
%\la{3.2}
%N_\ell (\zeta^{2}; 0) &= \l^\ell \prod_{(ij)}(\zeta^{2}-\cos\tau_{ij})  \longrightarrow 
%\notag \\
% N_\ell (\zeta^{2}; \theta_{0})   &=
%\l^\ell \prod_{(ij)}\big[\zeta^{2}(\cos^{2}\theta_{0}+\sin^{2}\theta_{0}\cos\tau_{ij})
%-\cos\tau_{ij}\big]\notag \\
%&=\big[ \l (1-\zeta^{2}\sin^{2}\theta_{0})\big]^{\ell}\,N_\ell\big(\frac{\zeta^{2}\cos^{2}\theta_{0}}
%{1-\zeta^{2}\sin^{2}\theta_{0}};  0\big).
%\end{align}
%
%
%
%Let us start with the 3-loop  expression for   $\langle W^{(\z)}\rangle_{\rm ladder}$ 
%containing  all (scalar and vector) ladder graph   contributions   to the $\z$-generalization of the circular WM loop. 
%Making the substitution in \rf{3.2}, i.e. replacing 
\be \la{3.15}
\l \ \  \to\ \  \l   ( 1 - \z_2^2) \ , \ \qquad \ \ \ \    \z^2 \ \  \to\ \   { \z_1^2  \ov 1- \z_2^2} 
\ . \ee 
 Keeping only the  contributions  with the  highest power of $\z_1$ and $\z_2$   at  each order in $\l$ 
  isolates   the  terms that   can only    come from  the  scalar ladder  graphs 
  and thus are  completely  determined  using  the replacement \rf{3.15}. %present computation.  
  Some discussion of the complete ladder contribution may be found in 
  Appendix~\ref{app:3loop2coupl}.
  We thus get the following  generalization of  the circular  loop    expression 
  \rf{2.16} ({\em i.e.} of the  case  when  $\z_1 =\z, \ \z_2=0$) 
\begin{align}
\la{3.16}
&\langle W^{(\z_{1},\z_{2})} \rangle_{\z\gg 1} = 1-\frac{1}{8}\lambda\, \z _2^2
+\lambda^{2}\,\Big[
\frac{\z_{2}^{4}}{192}+\frac{(\z_{1}^{2}+\z_{2}^{2})^{2}}{128\,\pi^{2}}
{-\frac{\z_{2}^{2}\,(\z_{1}^{2}+\z_{2}^{2})}{32\,\pi^{2}}\,\log\eps}
\Big]\notag \\
& +
\lambda^{3}\,\Big[
-\frac{\z_{2}^{6}}{9216}-\frac{\z_{2}^{2}(\z_{1}^{2}+\z_{2}^{2})^{2}}
{768\,\pi^{2}}-\frac{(\z_{1}^{2}+\z_{2}^{2})^{2}
(5\z_{1}^{2}+8\,\z_{2}^{2})}
{1536\,\pi^{4}} 
+\frac{5\,\z_{2}^{4}\,(\z_{1}^{2}+\z_{2}^{2})\,\zetaR(3)}{512\,\pi^{4}} \no \\
&\ \ {+\Big( \frac{\z_{2}^{4}\,(\z_{1}^{2}+\z_{2}^{2})}{384\,\pi^{2}} %\,\log\eps
+\frac{(\z_{1}^{2}+\z_{2}^{2})^{2}
(\z_{1}^{2}+2\,\z_{2}^{2})}{256\,\pi^{4}}\Big)
\log\eps} %\notag \\
%&\qquad 
 \ -\frac{\z_{2}^{2}(\z_{1}^{2}+\z_{2}^{2})^{2}}
{128\,\pi^{4}}\,\log^{2}\eps
\Big]+\dots.
\end{align}
In the special case of $\z_1=0$  (or $\theta_0={\pi \ov 2}$  in   \rf{1.4},\rf{1.5})  
corresponding to the $\z=\z_2$-deformation of the  
 $1\ov4$-supersymmetric loop     we find from \rf{3.15} 
\begin{align}
 &\langle W^{(0,\z_{2})} \rangle_{\z_{2}\gg 1} = 1-\frac{1}{8}\lambda\z _2^2\,
+\lambda^{2}\z_{2}^{4}\,\Big[
\frac{1}{192}+\frac{1}{128\,\pi^{2}}{-\frac{1}{32\,\pi^{2}}\,\log\eps}
\Big]
-\lambda^{3} \z_{2}^{6}\,\Big[
\frac{1}{9216}+\frac{1}{768\,\pi^{2}}\notag \\
&\qquad \qquad -\frac{5\,\zetaR(3)}{512\,\pi^{4}}
+\frac{1}{192\,\pi^{4}}{- \big(\frac{1}{128\,\pi^{4}} + \frac{1}{384\,\pi^{2}}\big) \log\eps
+\frac{1}{128\,\pi^{4}}\log^{2}\eps}
\Big]+\OO(\l^4) \ . \la{3.17}
\end{align}
This is equal to (\ref{2.13}) evaluated at $\z=0$ and  with 
$\lambda\to  - \lambda\,\z_{2}^{2}$    as required by \rf{3.15}
(taking  into account that  in \rf{3.15} we kept  only the highest powers of $\z_2$ at each order).

We may now  attempt to absorb the  divergences in \rf{3.15} by  a renormalization of the $\z_i$ 
couplings, {\em i.e.} by replacing $\z_i$ by their bare values 
\footnote{In a theory   with  dimensionless  running couplings  $g_i$   and loop counting parameter $\l$
the $\beta$-functions  have the form  $
\mu\frac{dg_{i}}{d\mu} = \beta_{i}(g) = \lambda\,\beta_{i}^{(1)}(g)
+\lambda^{2}\,\beta_{i}^{(2)}(g)+\lambda^{3}\,\beta_{i}^{(3)}(g)+\dots.$
In general,   the bare couplings $g_{i,\rm b}(\eps)$   depending on dimension-length  cutoff $\eps\to 0$ 
will be related  to renormalized  couplings  $g_{i}(\mu)$    depending on renormalization (dimension-mass) scale $\mu$
will have the general structure 
\begin{align}
\notag
g_{i,\rm b} &= g_{i}+\Big[ \lambda\,G_{1, i}^{(1)}(g)+ \lambda^{2}\,G_{1, i}^{(2)}(g)+\dots\Big]\,\log(\mu\,\eps) +\Big[
\lambda^{2}\,G_{2,i}^{(2)}(g)+
\lambda^{3}\,G_{2,i}^{(3)}(g)+\dots\Big]\,\log^{2}(\mu\,\eps)
+\dots.
\end{align}
The functions $G_{k, i}^{(\ell)}(g)$   can be expressed in terms  of $ \beta_{i}(g)$   and its derivatives  using the condition 
$ \mu {d g_{i,\rm b} \ov d \mu} =0$. % {\em i.e.}   the  coefficient of $\log\eps$ is just  $\beta_i (g)$, {\em etc.} 
For example,  in the case of one  coupling $g=g(\mu)$, one finds the familiar relation
\begin{align}
\notag
%\mu\frac{dg}{d\mu} &= \beta(g) = \lambda\,\beta_{1}(g)
%+\lambda^{2}\,\beta_{2}(g)+\lambda^{3}\,\beta_{3}(g)+\dots,\notag \\
g_{\rm b} &= g -  \beta(g) \log(\mu\,\eps)  + {1\ov 2} \Big[  \lambda^{2}\,\beta(g) \beta'(g) 
+\lambda^{3}\,
\big[\beta(g)\,\beta(g)\big]'+\dots\Big]\,\log^{2}\,(\mu\,\eps)\notag\\
&\ \ \ \ \ - \frac{1}{6} \Big[ \lambda^{3}\,\,\beta(g)\big[(\beta'(g))^{2}+\beta(g)\,\beta''(g)\big]
+\dots\Big]\,\log^{3}(\mu\,\eps)+\dots.\notag
\end{align}
}
%We now look for a bare-renormalized couplings redefinition suitable to our two couplings
%case, see the discussion in Appendix~(\ref{sec:RG}).
%Imposing the most general redefinition we find that cancellation of logarithms in 
%the above data is enough to fix the following form 
\begin{align}
\z_{1, \rm b} &= \z_{1}+\Big[\frac{\lambda}{8\pi^{2}}{F_1(\z_{1},\z_{2})}
+\dots\Big]\,\log\eps+\dots,\notag \\
\z_{2, \rm b} &= \z_{2}+\Big[
-\frac{\lambda}{8\pi^{2}}\,\z_{2}\,(\z_{1}^{2}+\z_{2}^{2}) %\notag \\ &
+\frac{\lambda^{2}}{(8\pi^{2})^{2}}
(\z_{1}^{2}+\z_{2}^{2})\Big(\frac{(\z_{1}^{2}+\z_{2}^{2})^{2}}{\z_{2}}
+\frac{\z_{1}}{\z_{2}}\,{F_1(\z_{1},\z_{2})}\Big)
+\dots
\Big]\,\log\eps\notag \\
&+\Big[
\frac{\lambda^{2}}{(8\pi^{2})^{2}}\,\Big(
-\frac{1}{2}\,\z_{2}\,(\z_{1}^{2}+\z_{2}^{2})\,(\z_{1}^{2}-3\,\z_{2}^{2})
-2\,\z_{1}\z_{2}\,{F_1(\z_{1},\z_{2})}
\Big)+\dots
\Big]\,\log^{2}\eps+\dots. \la{3.18}
\end{align}
For such redefinition to represent the solution of the RG equations  $\beta_i = \mu { d \z_i \ov d \mu} = - \frac{\lambda}{8\pi^{2}}
F_i(\z_{1},\z_{2}) + \OO(\l^2)$  we need to  require  that 
%Besides, we must impose that such a redefinition is consistent with the existence of 
%well-defined $\beta_{1,2}(\z_{1},\z_{2})$ functions. This determines the function $F(\z_{1},\z_{2})$
\be
F_1(\z_{1},\z_{2}) =  -\z_{1}\,(\z_{1}^{2}+\z_{2}^{2}) \ , \ \ \ \ \ \ \ \ \ \ \ 
F_2(\z_{1},\z_{2}) =  -\z_{2}\,(\z_{1}^{2}+\z_{2}^{2}) \ .\la{3.19}
\ee
This leads to the following expressions for the $\beta$-functions
\begin{align}
\beta_{1}(\z_{1},\z_{2}) &= \frac{\lambda}{8\pi^{2}}\,\z_{1}\,(\z_{1}^{2}+\z_{2}^{2})+\dots,\la{3.20}\\
\beta_{2}(\z_{1},\z_{2}) &= \frac{\lambda}{8\pi^{2}}
\z_{2}\,(\z_{1}^{2}+\z_{2}^{2})
-\frac{\lambda^{2}}{64\pi^{4}}\,
\z_{2}\,(\z_{1}^{2}+\z_{2}^{2})^{2}+\dots.\la{3.21}
\end{align}
For $\z\gg 1$ ({\em i.e.} $\z_1,\z_2 \gg 1$) the   one-loop part of these expressions is 
clearly consistent  with the previous result (\ref{3.13}).
 Inspired by \rf{3.21},\rf{3.13}  a natural  expectation for   the  structure of the  
 two-loop   $\beta$-functions  is 
\begin{align}\la{3.22}
\beta_{1}(\z_{1},\z_{2}) &=
\frac{\lambda}{8\pi^{2}}\,\z_{1}\,(\z_{1}^{2}+\z_{2}^{2}-1)
-\frac{\lambda^{2}}{64\pi^{4}}\,
\z_{1}\,(\z_{1}^{2}+\z_{2}^{2}-1)\,(\z_{1}^{2}+\z_{2}^{2}+\text{c}_{1})+\OO(\l^3),\notag\\
\beta_{2}(\z_{1},\z_{2}) &= \frac{\lambda}{8\pi^{2}}
\z_{2}\,(\z_{1}^{2}+\z_{2}^{2}-1)
-\frac{\lambda^{2}}{64\pi^{4}}\,
\z_{2}\,(\z_{1}^{2}+\z_{2}^{2}-1)\,(\z_{1}^{2}+\z_{2}^{2}+\text{c}_{2})+\OO(\l^3) \  , 
\end{align}
where c$_i$  are some constants to be determined. 
Then  the circle  $\z^2_1+\z^2_2 =1$ ({\em i.e.} the latitude WM loop  
with  $\z=1$ and arbitrary  $\theta_0$)  is still   a line of fixed points.
If it turns out that   $\text{c}_{1}\neq \text{c}_{2}$  then   for $\z\not=1$  the ratio $\z_1\ov \z_2$ or 
 $\theta_{0}$ starts running at two loops.
 
 %v2
 Note that if we formally set $\zeta_2=0$ in $\beta_1$ in \rf{3.22}  and  then  the coefficient of the  highest  power of 
 $\zeta_1=\zeta$ (cf. \rf{1.5})  in the 2-loop term will  suggest that the value of $b_2$ in \rf{2.21}   should be -1.

\section{Winding generalization of
deformed  $\frac{1}{4}$-BPS   WM loop}
\la{sec:wind}
%%%%%%%%%%%%%%%

Another generalization  of the latitude WM  loop is obtained by introducing  windings
of the $x^\m$ and  $n^m$  contours  in \rf{1.4}:  % To this aim, we replace in (\ref{1.1})
\be
x^{\mu}(\tau)\to x^{\mu}(\nu_{1}\,\tau),\qquad \qquad  n^{m}(\tau)\to n^{m}(\nu_{2}\,\tau),
\ee
where  $\nu_{1}, \nu_{2}$  are  integers. Let us denote by  
$\langle W^{(\z_{1},\z_{2})}(\nu_{1},\nu_{2}; \lambda)\rangle$ the
expectation value of the resulting $\z$-deformed  loop.
% for a generic choice of the deformation parameters. 
Local supersymmetry
requires $\z=1$ as a necessary condition, {\em i.e.} any loop with $n^m n_m =1$
is  locally supersymmetric
for generic values of  $\nu_{1}$ and $\nu_{2}$.
% \footnote{For a coupling $\int(i\,A_{\mu}\dot x^{\mu} + \Phi_{m}\,K^{m})$, the  condition
%for local supersymmetry is  $(\dot x^{\mu})^{2} = \bm{K^{2}}$.
%}
Global $1\ov 4$-supersymmetry  is present only if   $(\z_{1},\z_{2})=(0,1)$ ($\z=1, \theta_0={\pi\ov 2}$)  {\it and} $\nu_{1}=\nu_{2}$.
At one loop order  we have 
\begin{align}
&\langle W^{(\z_{1},\z_{2})}(\nu_{1},\nu_{2})\rangle = 
1+\lambda\,W_{1}^{(\z_{1},\z_{2})}(\nu_{1}, \nu_{2})+\mc O(\lambda^{2}),
\\
&W_{1}^{(\z_{1},\z_{2})}(\nu_{1}, \nu_{2}) = \frac{\nu_{1}^{2}}{8\,\pi^{2}}\int_{\tau_{1}>\tau_{2}} 
d\tau_{1}d\tau_{2}
\frac{\z_{1}^{2}+\z_{2}^{2}\cos(\nu_{2}\tau_{12})-\cos(\nu_{1}\tau_{12})}
{4\sin^{2}\frac{\nu_{1}\,\tau_{12}}{2}}.
\end{align}
Expanding in modes, we have\foot{The result  turns out to be finite so we do not need to 
introduce an exponential damping factor in the sum.}
\be
 W_{1}^{(\z_{1},\z_{2})}(\nu_{1}, \nu_{2}) %= \notag \\  
= \frac{\nu_{1}^{2}}{8\,\pi^{2}}\,\sum_{n=1}^{\infty}(-n)\,\int_{\tau_{1}>\tau_{2}} 
d\tau_{1}d\tau_{2} \, \Big[\z_{1}^{2}+\z_{2}^{2}\cos(\nu_{2}\tau_{12})-\cos(\nu_{1}\tau_{12})\Big]
\cos(n\,\nu_{1}\,\tau_{12}).
\ee
Assuming $\nu_{1}\neq \nu_{2}$, the integral is non-zero  for $n=1, \frac{\nu_{2}}{\nu_{1}}$.
The second contribution is present only  if $\frac{\nu_{2}}{\nu_{1}}\in \mathbb N$. 
Thus\footnote{Inspection of 
the special case $\nu_{1}=\nu_{2}$ reveals that the same expression is still  valid.}
\be
\la{4.5}
W_{1}^{(\z_{1},\z_{2})}(\nu_{1}, \nu_{2}) =
\begin{cases}
\frac{\nu_{1}^{2}}{8}, & \frac{\nu_{2}}{\nu_{1}}\not\in \mathbb N, \\
\frac{\nu_{1}^{2}}{8}\,\big(1-\z_{2}^{2}\,\frac{\nu_{2}}{\nu_{1}}\big), & \frac{\nu_{2}}{\nu_{1}}\in \mathbb N.
\end{cases}
\ee
This result can be cross checked with the more conventional dimensional regularization approach, 
see Appendix~\ref{app:dimreg}. Notice that (\ref{4.5})
 can be written in the form 
\be
W_{1}^{(\z_{1},\z_{2})}(\nu_{1}, \nu_{2}) = \nu_{1}^{2}\,F(\tfrac{\nu_{2}}{\nu_{1}}), \qquad
F(x) = \begin{cases} \frac{1}{8}, & x \text{ irrational} \\
\frac{1}{8}(1-\z_{2}^{2}\,x), & x \text{ rational}.
\end{cases}
\ee
%%%%%%%%%%%%%%%%%%%%%%%%
If $\z_1=0, \z_{2}=1$, the correction vanishes for $\nu_{1}=\nu_{2}$ which  corresponds  to 
multiply wound $1\ov 4$-supersymmetric loop. 
The opposite case of  $\z_1=1,\z_{2}=0$
 describes the multiply wound $\frac{1}{2}$-BPS WM loop
 (with no dependence on $\nu_{2}$):  here 
 the one-loop contribution is simply multiplied by $\nu_{1}^{2}$.
 This 
replacement rule $\lambda\to \lambda\,\nu_{1}^{2}$ extends to all orders as follows from the 
matrix model solution \cite{Drukker:2000rr,Drukker:2005kx}.\footnote{
Let us add a   technical remark. For $\nu_{1}\neq 1$, the treatment of the planar restriction on diagrams
becomes  quite involved at higher loops. The reason is that we are computing integrals  like 
$
\int_{0}^{2\pi} d\tau_{1}\dots d\tau_{n} \,\langle \Phi(x(\tau_{1}))\dots \Phi(x(\tau_{n}))\rangle_{
\rm planar}
$ and, if $x(\tau)$ describes a multiply wound circle, %as is the case when $\nu_{1}>1$, 
then the domain of 
integration has to be properly  split in order to select the correct planar contractions. A detailed 
discussion of this issue can be found in  \cite{Bianchi:2016gpg}.}

In the remaining part of this section, we shall restrict consideration to the case 
of  %(the sign of $\nu$  is not relevant; we shall assume $\nu >0$)   
\be  \nu_{1}=1\ , \qquad  
\nu_{2}\equiv \nu\in \mathbb N^+ \ , \qquad \qquad   (\z_{1},\z_{2}) = (0, \z) \ .  \la{47} \ee
% \footnote{Sign of $\nu$ is not relevant. We shall conventionally assume 
%$\nu>0$.}
%\be
%n^{1}=\sin\theta_{0}\,\cos(\nu\tau),\qquad 
%n^{2}=\sin\theta_{0}\,\sin(\nu\tau),\qquad 
%n^{3}=\cos\theta_{0}, \qquad \nu=1,2,3,\dots.
%\ee
The choice of 
 $(\z_{1},\z_{2}) = (0, \z)$   corresponds to  1-parameter 
 deformation of the $1\ov 4$-supersymmetric  WM loop  \ci{Zarembo:2002an}
 (with trivial expectation value for $\z=1$). 
According to (\ref{4.5}), at one-loop  order we get 
%we have  simply the replacement
\be
\la{4.8} \langle W^{(0,\z)}(1,\nu)\rangle = 
1+
 \frac{\lambda}{8}(1-\z^{2}\,\nu)  + \mc O(\l^2)  \ .
\ee
In    the  winding   generalization of the 
supersymmetric  case  ($\z=1$) we get  a  non-trivial 
1-loop  contribution $\frac{\lambda}{8}\,(1-\nu)$:  %  which 
%This vanishes for $\nu=1$ but not for $\nu>1$. This is because 
%at $\z=1$, the winding numbers 
the   choice of  unequal   winding  numbers 
$\nu_{1}=1$ and $\nu_{2}=\nu>1$ breaks space-time supersymmetry and  thus 
$
\mc W(\nu) \equiv  \langle W^{(0,1)}(1,\nu)\rangle
$
is  then a non-trivial function of $\nu$ and $\lambda$. 

To determine what happens at 
two loops, we  shall first  consider the  ladder diagrams and then  include  all other contributions  indirectly using information 
from  defect CFT$_{1}$  corresponding to perturbations 
  near  the $\z=0$  conformal point.
 
 \def \G  {\Gamma} 
 
%%%%%%%%%%%%%%%%%%%%%%%%%%%
\subsection{Two-loop ladder contribution}

The two ladder diagrams contributing at two loops are 
(14)(23) and (12)(34). From the results of 
 Appendix~\ref{app:winding}, we obtain (here $\z$ is the bare coupling)
\begin{align}
\la{4.10}
\langle  W^{(0, \z)}&(1, \nu)\rangle_{\rm ladder} = 1+
\frac{\lambda}{8}(1-\z^{2}\,\nu)+\lambda^{2}\,\Big[
\frac{(1-\z^{2}\nu)^{2}}{192}+\frac{1+\z^{2}(\z^{2}-2)(2\nu-1)}{128\,\pi^{2}}\notag \\
&+\frac{\z^{2}(1+\nu-\z^{2}\nu)}{32\,\pi^{2}}\big[\psi(\nu)-\psi(1)\big]
+\frac{\z^{2}\nu(2-\z^{2}\nu)}{64\,\pi^{2}}\,\big[\psi^{\prime}(\nu)-\psi^{\prime}(1)\big]\notag \\
&{-\frac{(\z^{2}-1)(\z^{2}\nu-1)}{32\,\pi^{2}}\,\log\eps}
\Big]+\mc O(\lambda^{3}).
\end{align}
where $\z$ stands for bare coupling and  $\psi(\nu)= \G'(\nu)/\G(\nu)$.
% is the Euler $\psi$-function. 

The two-loop terms  $\sim\lambda^{2}\z^{0}$ and $\sim\lambda^{2}\,\z^{2}$
 potentially   receive  contributions  from   non-ladder diagrams too.
 Instead, the terms  $\sim \lambda^{2}\,\z^{4}$ come
only from ladder diagrams. A special  limit  where non-ladder diagrams may be 
neglected is that of  $\z\gg 1$  with $\l\, \z^2$=fixed.
%In this case, 
%only ladders are important with non-ladder diagrams being sub-dominant.
The coefficient of the UV logarithm is linear in $\lambda^{2}\z^{4}\nu$ and is absorbed by 
a shift in the one-loop term with 
$\beta\sim \lambda\z^{3}$  having  no dependence on $\nu$, i.e. 
being the same as  the previous  one  for $\nu=1$.
%This shift removes the divergent term $\sim\log\eps$ and admits an arbitrary finite contribution
Introducing  the  renormalization scale $\muMR$ (corresponding to the mode regularization scheme, see discussion after (\ref{2.15}))  we find for the {\it renormalized} expression\foot{
Another interesting
regime is $\nu\gg 1$  where  we have from (\ref{4.10})
\be
\notag 
\langle W^{(0, \z)}(1,\nu)\rangle_{\rm ladder} \stackrel{\nu\gg 1}{=} 
1-\frac{\lambda}{8}\z^{2}\nu + \frac{\lambda^{2}}{128}\,\z^{4}\nu^{2}+\mc O(\lambda^{n}\nu^{n-1}).
\ee
%This is non-zero and therefore 
Possible additional contributions from non-ladder diagrams
 involve smaller powers of $\z$ at each loop. }
(here $\z$ is    the 
renormalized coupling)
\begin{align}
\la{4.11}
\langle  W^{(0, \z)}(1, \nu)\rangle & \stackrel{\z\gg 1}{=} 1-\frac{\lambda}{8}\,\z^{2}\,\nu+
\lambda^{2}\,\z^{4}\,\Big[
\frac{\nu^{2}}{192}+\frac{2\,\nu-1}{128\,\pi^{2}}\notag \\
&
-\frac{\nu\,[\psi(\nu)-\psi(1)]}{32\,\pi^{2}}
-\frac{\nu^{2}\,[\psi'(\nu)-\psi'(1)]}{64\pi^{2}}
{-\frac{\nu}{32\,\pi^{2}}\,\log(\muMR\,R)}
\Big]+\dots.
\end{align}
At $\z=1$, the two-loop extension of (\ref{4.8}) due  to ladder diagrams only   reads
\be
\la{4.12}
\mc W(\nu)_{\rm ladder} = 
\langle W^{(0, 1)}(1, \nu)\rangle_{\rm ladder} = 1+\frac{\lambda}{8}\,(1-\nu)+\lambda^{2}\,W_{2}(\nu)
+\mc O(\lambda^{3}),
\ee
where 
$W_{2}(\nu)$ is the following non-trivial function  %of the winding $\nu$ 
\footnote{
The large $\nu$ expansion of (\ref{4.13}) is
\be
\notag
W_2(\nu) = \frac{\lambda^{2}}{128}\,\Big[
\nu ^2-2\,\Big(1+\frac{2}{\pi ^2}\Big) \nu +
\frac{4\, (\log \nu+\gamma_{\rm E}) }{\pi ^2}+\frac{5}{\pi^2}+\frac{2}{3}+\mc O(\nu^{-1})\Big].
\ee
}
\begin{align}
\la{4.13}
W_2(\nu) \te= \frac{1}{64\,\pi^{2}}\Big[
(1-\nu)+{\pi^{2}\ov 3}\,(1-\nu)^{2}
+2[\psi(\nu)-\psi(1)]+\nu\,(2-\nu)\,
[\psi^{\prime}(\nu)-\psi^{\prime}(1)]
\Big]
\end{align}
Its explicit  values for $\nu=1,2,3,\dots$ are 
$
W_2(\nu) = 0, \ \tfrac{1}{192}+\tfrac{1}{64\pi^{2}}, \ 
\tfrac{1}{48}+\tfrac{19}{256\,\pi^{2}}, \dots
$

%%%%%%%%%%%%%%%%%
\subsection{Extracting  non-ladder contribution from  $\z=0$  defect CFT$_{1}$}

In (\ref{3.14}), we have obtained the complete two-loop  contribution to $\langle W^{(\z_{1},\z_{2})}
\rangle$  {without winding} 
evaluated in dimensional regularization. The scheme dependence is contained  in the constant 
$a=5+4\,\log(\muDR\,R)$,   where 
 $\muDR$  corresponds to  dimensional regularization. 
 The extension to  a non trivial winding $(\nu_{1}, \nu_{2}) = (1,\nu)$
 would require the evaluation from scratch of all  
 (ladder, self-energy and 
internal vertex)   two-loop diagrams  within the same scheme.
% \footnote{A natural candidate being dimensional
%regularization, because mode regularization is quite simple for ladder diagrams, but would require
%an extension to deal with the other diagrams, in particular those with internal vertices.}

Nevertheless, there is a shortcut that bypasses the actual evaluation of the non-ladder diagrams.
In general,  $\langle W^{(0,\z)}(1,\nu)\rangle$ has $\z^{0}$, $\z^{2}$, and $\z^{4}$ contributions.
The $\z^{0}$ contribution  can not depend on the $\nu_2$-winding (as for $\z=0$  the scalar coupling in  \rf{1.44}  vanishes) 
and thus can be computed at $\nu=1$. 
The 
$\z^{4}$ contribution comes only from ladder diagrams  and may be found  using  mode regularization,
{\em i.e.}  from  (\ref{4.11}).
Finally,
the $\z^{2}$ contribution can be extracted by a $\z=0$ defect CFT$_{1}$ 
 where 
all the necessary 
 data does not depend on $\nu$ and can be fixed by comparison with the 
complete result (\ref{3.14}) evaluated in dimensional regularization. 
As a last step, the relation between  the 
  dimensional regularization  and the mode regularization  schemes,  {\em i.e.}
  between $\muDR$ and $\muMR$
can be  found   via a suitable analysis of finite renormalizations in the two schemes.

\subsubsection*{Using  CFT data}

We begin with the detailed determination of  the $\z^{2}$ contribution.
Starting   with 
the winding  generalization of the   scalar  coupling  in \rf{1.44}
\be
\la{4.15}
\z_{1}\,\Phi^{3}+\z_{2}\,[\cos(\nu\tau)\,\Phi^{1}+\sin(\nu\tau)\,\Phi^{2}] \ , 
\ee
we  may view  the expansion in powers of $\z_1, \z_2$  as   perturbation theory near the 
  conformal point $\z_1=\z_2=0$
%We consider a $(\z_{1},\z_{2})=(0,\z)$ perturbation with respect to the standard 
%non supersymmetric conformal point
$\z_{1}=\z_{2}=0$ where all 6 scalar fields 
$\{\Phi^{m}\}$ are equivalent \ci{Beccaria:2017rbe}.
The scalars  restricted to the  circular loop   correspond to 
operators    in 1d  CFT  (which we will also denote as $\Phi$). 
In  the products  of couplings   and operators 
may  involve the bare or renormalized quantities,  
$ %\be
\z_{\rm b}\,\Phi_{\rm b} = \z_{\rm r}\,\Phi_{\rm r}  . $
%\ee
%by definition. For our application, 
Here it  is convenient to work with the renormalized  operators and couplings
(suppressing the label   ``r''). In particular, the two-point function of the 
 CFT$_{1}$  operators   corresponding  to the scalar fields
 restricted to the circular  loop of radius $R$   and  renormalized at scale $\muDR$, 
reads  (here $ |s_{12}| = |x(\tau_1) -x(\tau_2)| = 2R\,\sin\frac{\tau_{12}}{2}$) % is the chordal distance)
\begin{align}
& \llangle \Phi^{m}(\tau_{1})\Phi^{n}(\tau_{2})\rrangle = \delta^{mn}\,\frac{\muDR^{2}\,
C_{0}(\lambda)}{
|2\,\muDR\,R\,\sin\frac{\tau_{12}}{2}|^{2\,\Delta}} = 
\frac{\delta^{mn}}{|s_{12}|^{2}}\,\,\frac{C_{0}(\lambda)}{
|\muDR\, s_{12}|^{2\,(\Delta-1)}} 
, \la{417} \\
& \Delta = 1-\frac{\lambda}{8\,\pi^{2}}+\mc O(\lambda^{2}), \qquad\qquad 
C_{0}(\lambda) = \frac{\lambda}{8\pi^{2}}-C_{0}^{(1)}\,\frac{\lambda^{2}}{\pi^{2}}
+\mc O(\lambda^{3}) \ . \la{4.17}
\end{align}
Expanding the  expectation value $\langle W^{(0,\z)}(1,\nu)\rangle$ at small $\z$ gives
\begin{align}
\la{4.18}
\log \langle W^{(0,\z)}(1,\nu)\rangle &= \log \langle W^{(0,0)} \rangle + \z^{2}\,\mc D_{2}(\nu) + 
\mc O(\z^{4}) \ , 
\end{align}
where the  first term is independent on $\nu$ %(there is no scalar coupling at $\z=0$) and has been 
and was  already computed in \cite{Beccaria:2017rbe} at two loops
 %It   is   given by the following  finite  %scheme independent 
%expression
\be\la{4417}
\langle W^{(0,0)}\rangle = 1+\frac{\lambda}{8}+\lambda^{2}\,\Big(\frac{1}{192}+\frac{1}{128\,\pi^{2}}\Big)+\mc O(\lambda^{3}).
\ee
The second term in (\ref{4.18}) is  expressed in terms of the  CFT$_{1}$ 2-point function
\be\la{4.19} 
\mc D_{2}(\nu) = \ha 
\int_{0}^{2\pi}d\tau_{1}\,d\tau_{2}\,\llangle \widetilde \Phi(\tau_{1})\,
\widetilde\Phi(\tau_{2})
\rrangle,\qquad \widetilde\Phi(\tau) = \cos(\nu\tau)\,\Phi^{1}+\sin(\nu\tau)\,\Phi^{2}.
\ee
Using (\ref{4.17}), we can write  it as follows \footnote{We omit the trivial overall
$1/R^{2}$ factor,  but  will  keep an explicit $R$ in the $\log(\muDR R)$ below.}
\begin{align}
\mc D_{2}(\nu) &= \ha \muDR^{-2\,(\Delta-1)}{C_{0}(\lambda)}\int_{0}^{2\pi}d\tau_{1}d\tau_{2}
\frac{\cos(\nu\tau_{1})\cos(\nu\tau_{2})+\sin(\nu\tau_{1})\sin(\nu\tau_{2})}{(
4\sin^{2}\frac{\tau_1-\tau_2}{2})^{\Delta(\lambda)}} \notag \\
&= \muDR^{-2\,(\Delta-1)}\,2^{-2\,\Delta(\lambda)}\,\pi\,C_{0}(\lambda)\,
\int_{0}^{2\pi}d\tau\,\cos(\nu\tau)\,\left(\sin^{2}\frac{\tau}{2}\right)^{-\Delta(\lambda)}.
\end{align}
This integral can be computed using  the method discussed in
 Appendix~\ref{app:dimreg}\footnote{Notice that the second order correction to $\Delta$ does not appear at this order 
since at weak coupling $C_{0}(\lambda)$ starts at order $\lambda$.}
\be
\la{4.22}
\mc D_{2}(\nu) = -\frac{\lambda}{8}\,\nu
+\lambda^{2}\,\Big[
-\frac{\nu\,\log(\muDR\,R)}{32\,\pi^{2}}
-\nu\,C_0^{(1)}+
\frac{1-2\,\nu+2\,\nu\,[\psi(\nu)-\psi(1)]}
{64\,\pi^{2}}\Big]+\mc O(\lambda^{3}).
\ee
To determine $C_{0}^{(1)}$, {\em i.e.} the second order correction to the 2-point function normalization in \rf{4.17},
let us   specialize (\ref{4.22}) to  $\nu=1$  
\be
\la{4.23}
\mc D_{2}(1) = -\frac{\lambda}{8}
+\lambda^{2}\,\Big[
-\frac{\log(\muDR\,R)}{32\,\pi^{2}}-C_0^{(1)}-\frac{1}{64\,\pi^{2}}\Big]+\mc O(\lambda^{3}).
\ee
This  can be compared to  the coefficient of $\z_{2}^{2}$ 
in the expansion of $\log \langle W^{(0,\z_2)}\rangle$, {\em i.e.} using the DR 
result (\ref{3.14}) % ($\z_2\equiv \z$) 
\be
\la{4.24}
\log\langle W^{(0,\z_2)}\rangle = \log\langle W^{(0,0)}\rangle +
\z_2^{2}\,\Big[
-\frac{\lambda}{8}+\lambda^{2}\,\Big(
\frac{1}{192}-\frac{3}{64\,\pi^{2}}-\frac{\log(\muDR\,R)}{32\,\pi^{2}}
\Big)+\mc O(\lambda^{3})
\Big]
+\mc O(\z_2^{4})
\ee
Comparing  (\ref{4.23}) and (\ref{4.24}) gives 
\be
\la{4.25}
C_{0}^{(1)} = -\frac{1}{192}+\frac{1}{32\,\pi^{2}}.
\ee

\subsubsection*{Matching  regularization schemes}

Next, we  need to  finding a relation
between the mode regularization scale $\muMR$ appearing in (\ref{4.11}), 
and the dimensional regularization scale $\muDR$
appearing in (\ref{4.22}). This can be achieved by comparing the $\z_{2}^{4}$ term of the
DR expression (\ref{3.14}) at $\z_{1}=0$  
with the $\nu=1$ values of the MR result (\ref{4.11}). One gets
\be
\la{4.26}
\log(\muDR\,R)+\log(\muMR\,R)+1=0.
\ee

\subsubsection*{Reconstruction of full $\langle W^{(0,\z)}(1,\nu)\rangle$}

As a final step,  collecting  all terms  that make up the full expression of
$\langle W^{(0,\z)}(1,\nu)\rangle$  and exponentiating (\ref{4.18}) we obtain 
\begin{align}
\langle & W^{(0,\z)}(1,\nu)\rangle 
=\z^{0}\Big[1+\frac{\lambda}{8}+\Big(\frac{1}{192}
+\frac{1}{128\,\pi^{2}}\Big)+\mc O(\lambda^{3})\Big]\notag \\
&+\z^{2}\Big\{- \frac{\lambda}{8}\nu -\frac{\lambda^{2}}{64}
+\lambda^{2}\,\Big[
{-\frac{\nu\log(\muDR\,R)}{32\,\pi^{2}}}
-\nu\,C_0^{(1)}+
\frac{1-2\,\nu+2\,\nu\,[\psi(\nu)-\psi(1)]}
{64\,\pi^{2}}\Big]+\mc O(\lambda^{3})
\Big\}\notag\\
&+\z^{4}
\Big\{
\lambda^{2}\Big[
\frac{\nu^{2}}{192}+\frac{2\,\nu-1}{128\,\pi^{2}}
-\frac{\nu\,[\psi(\nu)-\psi(1)]}{32\,\pi^{2}}
-\frac{\nu^{2}\,[\psi'(\nu)-\psi'(1)]}{64\pi^{2}} \no \\
&\ \   \qquad  \qquad \qquad {-\frac{\nu}{32\,\pi^{2}}\,\log(\muMR\,R)}
\Big]+\mc O(\lambda^{3})
\Big\}
\end{align}
where we have taken the $\z^{4}$ term from the mode regularization calculation (\ref{4.11}).
 After
some regrouping, and replacing $\muMR$ by  $\muDR$
using  (\ref{4.26}), we  finally obtain 
\begin{align}
\la{4.28}
\langle  W^{(0, \z)}(1,\nu)  \rangle = 
1 &+\frac{\lambda}{8}\,(1-\z^{2}\,\nu)\notag \\
&+\lambda^{2}\,\Big[
\frac{1-3\,\z^{2}}{192}+\frac{\z^{2}\,\nu\,(1+\z^{2}\,\nu)}{192}
-\frac{\z^{4}-2\z^{2}-1}{128\,\pi^{2}}+\frac{\z^{2}(3\,\z^{2}-4)\,\nu}{64\pi^{2}}\notag \\
&\quad \qquad  -\frac{\,\nu\,  \z^{2}\,(\z^{2}-1)\,[\psi(\nu)-\psi(1)]}{32\,\pi^{2}}
 -\frac{\z^{4}\,\nu^{2}[\psi'(\nu)-\psi'(1)]}{64\,\pi^{2}}\notag \\
&\quad \qquad +{\frac{\nu\,\z^{2}\,(\z^{2}-1)}{32\,\pi^{2}}\,\log(\muDR R)}
\Big]+\mc O(\lambda^{3})\ . 
\end{align}

%%%%%%%%%%%%%%%%%%%%%%%%%%%
\subsection{Winding deformation of the $\frac{1}{4}$-supersymmetric loop}

Setting $\z=1$  gives  the expectation value of the $\frac{1}{4}$-supersymmetric WM 
loop   generalized to  winding   number 
 $\n$  along the big circle of $S^5$ ($\theta_0 ={\pi\ov 2}$).
 For $\n>1$   global supersymmetry is broken,   so we  expect to get a non-trivial function of 
 $\l$.
We   find from  (\ref{4.28})  the following  finite two-loop expression 
\begin{align}
\la{4.29}
\mc W(\nu) \equiv  \langle W^{(0,1)}(1,\nu)\rangle= 1&+ \frac{\lambda}{8}(1-\nu)+\lambda^{2}\Big[-
\frac{(1-\nu)(\nu+2)}{192}-\frac{1-\nu}{64\,\pi^{2}}\no \\
&\qquad \qquad\qquad  \qquad  -\frac{\nu^{2}[\psi'(\nu)-\psi'(1)]}
{64\,\pi^{2}}
\Big]+\mc O(\lambda^{3}).
\end{align}
Then  $\mc W(0)$ is  the standard Wilson loop expectation value \rf{4417},
$\mc W(1)=1$  while, e.g.,     for $\n=2$ and 3    we get \begin{align}
\mc W(2) &= 1-\frac{\lambda}{8}+\lambda^{2}\,\Big(\frac{1}{48}+\frac{3}{64\,\pi^{2}}\Big)
+\mc O(\lambda^{3}),\quad
\mc W(3) = 1-\frac{\lambda}{4}+\lambda^{2}\,\Big(\frac{5}{96}+\frac{37}{256\,\pi^{2}}\Big)
+\mc O(\lambda^{3}).
\end{align}
\iffa 
As we mentioned previously, even if we are considering 
$\z=1$, the winding numbers 
$\nu_{1}=1$ and $\nu_{2}=\nu>1$ break space-time supersymmetry and this is why we can get
a non-trivial function of $\nu$
and $\lambda$. Two-loop finiteness shows that this kind of  supersymmetry breaking 
is rather mild due to the preserved local supersymmetry. 
\fi 
%Loosely speaking, this is similar to the 
%spontaneous breaking of supersymmetry due to winding by Scherk-Schwarz 
%mechanism \cite{Sxxxcherk:1978ta}
%discussed in phenomenological applications \cite{Dxxxvali:1996bg,Kxxxaplan:2001cg}.
Since this case ($\z=1$)    corresponds to locally supersymmetric WM loop 
in finite $\N=4$ SYM theory, the   function   $\mc W(\nu)$  should  be finite at   all orders
of expansion in $\l$.
% This is basically
%because $\lambda$ does not renormalize in $\mc N=4$ SYM and $\nu$ is a fixed quantized (integer)
%parameter. Thus,
It   would be very interesting to compute it exactly (possibly using integrability). 
  To recall,   for the  circular WM loop (corresponding to contour 
  %v2
  in $S^5$ shrinking to a point, or $\theta_0=0$) % which is trivial in $S^{5}$,
the   winding of the space-time  circle 
can be completely absorbed into  a  rescaling of $\lambda$ (as seen, {\em e.g.}, 
 from  the matrix model representation   \cite{Drukker:2000rr}). On the other hand, the unwound $1\ov 4$-supersymmetric loop is  trivial, $\mc W(1)=1$ \cite{Zarembo:2002an}.
It would be very interesting if $\mc W(\nu)$  could  be  reproduced   by a matrix model   leading to the expansion (\ref{4.29}). At the same time, 
in the absence   of global supersymmetry a localization to  a matrix model may  seem to be 
unlikely.

\def \s  {\sigma}\def \sql {\sqrt\lambda}

Another   interesting   question  is  to  try to  understand  the     behaviour  of \rf{4.29} 
  at strong coupling  using $AdS_5 \times S^5$ string theory   picture.
   At leading order in inverse  string  tension,  one may generalize the discussion 
  in  
\cite{Zarembo:2002an} to any  winding  $\nu$  in  the big circle of $S^5$. 
In this case the  induced  world-sheet  metric is (here $\tau \in (0, 2\pi)$, $\s \in (0, \infty)$)
\be \la{111}
ds^2 =  \Big[ {1 \ov \sinh^2 \s}   + {\nu^2\ov \cosh^2(\nu \s)} \Big] \big( d\tau^2  + d\s^2\big) \ , 
\ee
interpolating  between the $AdS_2$    metric  (for $\nu=0$,   corresponding to  the standard  circular loop in $AdS_5$)   and  the $\nu=1$ case    when the   world sheet   ends 
   also  on a  big  circle of $S^5$   \cite{Zarembo:2002an}.
   The  string action   proportional  to  the  regularized area is 
    then  ($T= {\sqrt \l \ov 2 \pi}$)\foot{In general, in the case of two
    non-trivial  windings $\n_1$ and $\n_2$  the  euclidean solution  in conformal gauge  has  the   $AdS_5$ part   (in Poincare coordinates) as 
    $x^\m= {1 \ov \cosh ( \n_1 \s)} \{ \cos (\n_1 \tau), \sin (\n_1 \tau) \}, \ z=\tanh (\n_1 \s)$
    and  the $S^5$ part  as  \ $\varphi= \nu_2 \tau, \  \cos \theta=\tanh (\nu_2 \s)$. Then  the induced metric is 
    $ds^2 =  \big[ {\n^2_1 \ov \sinh^2 (\n_1\s)}   + {\nu_2^2\ov \cosh^2(\nu_2 \s)} \big] \big( d\tau^2  + d\s^2\big) $.
    The resulting  finite part of the string action is 
    $I_{\rm fin} = - \sql\, (|\nu_1|-|\nu_2|)$   which vanishes in the supersymmetric case 
    $\n_1=\nu_2$ (cf. \ci{Drukker:2007qr}).}
\be  
I=  2\,\pi\, T\, \int_{\eps}^{\infty }d\s \,  \Big[ {1 \ov \sinh^2 \s}   + {\nu^2\ov \cosh^2(\nu \s)} \Big]  
= {\sql \ov \eps} - \sql(1-|\nu|)+ \mc O(\eps) \ . \ee
The  renormalized   action    leads to the 
 following prediction for the   behaviour   of % Thus   one should expect that 
\rf{4.29}    at strong coupling:
\be  
\mc W(\nu)   \stackrel{\l\gg 1}{=}\   e^{  \sql\, (1-|\nu|)} \ . \la{666}
\ee

%%%%%%%%%%%%%%%%%%%%%%%%%%%%%%%%%
\section*{Acknowledgments}
We are grateful to   S. Giombi and K. Zarembo for useful discussions of related questions.
%The work of S.G. is supported in part by the US NSF under Grant No.~PHY-1620542.
AAT  was  supported by STFC grant ST/P000762/1  
and  by   the Russian Science Foundation grant 14-42-00047 at Lebedev Institute.

%%%%%%%%%%%%%%%%%%

\appendix
\section{Calculation  of three-loop ladder diagrams in mode regularization}
%to three loops}
\la{sec:3L}

%\bf Remark:} 

Below  we  will  summarize  some details of computation of higher loop  ladder graphs
in the case of the circular Wilson loop. 
These diagrams   correspond to  "rainbow"  diagrams in the case of a  straight line
and we shall plot them  this way below for simplicity. 
By "scalar ladders" we  will  mean the ladder diagrams  with only scalar propagators  %contribution from    %to a ladder diagram from 
%scalar exchange 
(i.e.  coming from   the first term in $\mathcal G(\tau)$  in \rf{2.2}); they  give    the only relevant contributions in the large $\z$ limit.

\subsection{Two-loop scalar  ladders }
\la{sec:prelim}

Let us  first  recall  a   simple computation of 2-loop scalar  ladders  from  \ci{Beccaria:2017rbe}. 
Let us consider the $\zeta^{2}$ term in the $(14)(23)$ ladder diagram (for notation see \rf{2.3}) 
\be
I = \int [d^{4} \bm{\tau}] \,
\mathscr D(\tau_{14})\,\mathscr D(\tau_{23}),\qquad\qquad 
\mathscr D(\tau) = \frac{1}{4\sin^{2}\frac{\tau}{2}}.
\ee
Using the  regularized expression for   $\mathscr D(\tau)$ in   \rf{2.4}  we get  (following  the notation in \rf{2.7}) 
\begin{align}
I^{(2)}_{(14)(23)}  &= \sum_{n,m=1}^{\infty} e^{-\eps\,(n+m) }\, S_{n,m},\qquad %\text{with}\quad
S_{n,m} =  n m \int  [d^{4}\bm{\tau}]
\cos(n\,\tau_{14})\cos(m\,\tau_{23}).
\end{align}
After integration, we find
\begin{align}
S_{n,m} 
= &\frac{1}{2 m n (m^2-n^2)^2}\Big[2 m^2 (m^2-3 n^2)-2 (m^2-n^2)^2 \cos (2 \pi  n)\notag \\
&+n^2 (m+n)^2  \cos (2 \pi  (m-n))+n^2 (m-n)^2 \cos (2 \pi  (m+n))\Big].
\end{align}
For generic integer $n,m$ one checks that $S_{n,m}=0$. The 
removable singularity occurring for  $n=m$ must be considered separately.
In this case, one finds     $S_{n,n}=-\pi^{2}$. Thus
\be
I^{(2)}_{(14)(23)} = -\frac{1}{2} \pi ^2 (\coth \eps -1) = 
-\frac{\pi ^2}{2 \eps }+\frac{\pi ^2}{2}+\mc O(\eps).\la{A.4}
\ee
Dropping the  power divergence (which is equivalent to  $\zeta$-function regularization) 
   we reproduce the  coefficient of the $\z^4$ term  in  $I_{(14)(23)}$ \rf{2.10}
   (other terms in  \rf{2.10}   get contributions also from ladder diagrams  with vector  propagators). 
   
%The finite part is what we used: $I = \pi^{2}/2$. In this case, exponential regularization is same as 
%$\zeta$-function regularization.

\subsection{Three-loop scalar  ladders}
\la{appA2}

%$\bullet$  

\subsection*{$\bullet$ \ (16)(25)(34)}

\begin{figure}[!htb]
\begin{center}
\includegraphics[scale=0.45]{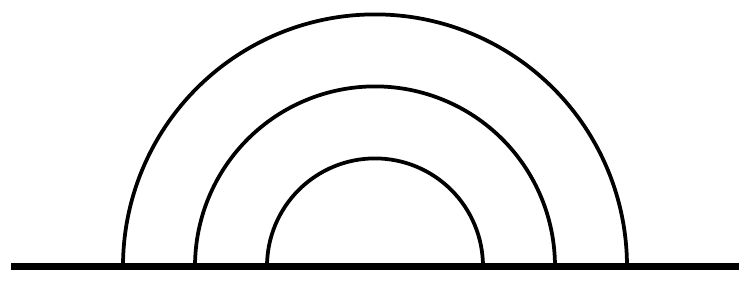}\qquad \qquad
\caption{
Ladder diagram associated with the (16)(25)(34) contraction. Here and in the following pictures, 
the horizontal thick line represents the loop parametrized by $\tau\in[0,2\pi]$. Scalar propagators
are half-circles attached to the loop.
\la{fig:16-25-34} }
\end{center}
\end{figure}

At three loops we have similarly 
\be
I^{(3)}_{(16)(25)(34)} = \int[d^{6}\bm{\tau}]\,
\mathscr D(\tau_{16})\,\mathscr D(\tau_{25})\,\mathscr D(\tau_{34}).
\ee
Using again (\ref{2.4}), we find 
\begin{align}
I _{(16)(25)(34)}&= \sum_{n,m,s=1}^{\infty} e^{-\eps(n+m+s) }\ S_{n,m,s}\ ,\la{A.6}  \\ 
S_{n,m,s} &= -n\,m\,s\,\int [d^{6}\bm{\tau}]\,
\cos(n\,\tau_{16})\cos(m\,\tau_{25})\cos(s\,\tau_{34})\ . \la{A.7}
\end{align}
The integral  in \rf{A.7}  is non zero in the following cases
\be
\la{A.8}
S_{n,m,s} = \begin{cases}
\frac{\pi^{2}}{s}, & m=n, \ m\neq s,\ s \neq 2\,m, \\
-\frac{\pi^{2}}{n}, & m=s, \ m\neq n,\ n\neq 2\,m \\
-\frac{m\,\pi^{2}}{2\,n\,(m+n)}, & s=m+n,\ m\neq n,\\
\frac{\pi^{2}}{4m}, & s=2m,\ m=n,\\
-\frac{m\,\pi^{2}}{2\,n\,(m-n)}, & m=s+n, \\
\frac{m\,\pi^{2}}{2\,n\,(m-n)}, & n=m+s,\ m\neq s, \\
-\frac{3\,\pi^{2}}{4\,m}, & n=2m,\ m=s, \\
0, & \text{else}.
\end{cases}
\ee
%We can try to regularize the sum over $n,m,s$ by a symmetric exponential $e^{-\eps(n+m+s)}$.
Due to symmetry of the sum in \rf{A.6}  the  contributions  of the first two cases   in (\ref{A.8}) mutually cancel. 
The other  cases   give
%\foot{Let us note  that one can compute  the sum of all contributions with fixed $k=n+m+s$. One finds
%\be
%\left. S_{n,m,s}\right|_{n+m+s=k} = -\pi^{2}\frac{1+(-1)^{k}}{2}\,\Big[
%-1+2\,\gamma_{\rm E}+\frac{2}{k}+2\,\psi\left(\frac{n}{2}\right)\Big], \no 
%\ee
%where $\psi(z)$ is the digamma function. The sum of this expression times 
%$\exp(-k\,\eps)$ is recognized by Mathematica in closed form (and of course agrees with (\ref{A.9})).
%}
\begin{align}
\la{A.9}
& I^{(3)}_{(16)(25)(34)}(\eps) =  -\frac{\pi^{2}}{2}\sum_{n,m=1}^{\infty}e^{-2\eps\,(n+m)}\frac{m}{n\,(n+m)}
+\sum_{m=1}^{\infty}e^{-4\,\eps\,m}\Big(\frac{\pi^{2}}{4m}
+\frac{\pi^{2}}{4m}
\Big)\notag \\
& -\frac{\pi^{2}}{2}\sum_{n,s=1}^{\infty}e^{-2\eps\,(n+s)}\frac{s+n}{n\,s}
-\frac{\pi^{2}}{2}\sum_{m,s=1}^{\infty}e^{-2\eps\,(m+s)}\frac{m}{(m+s)\,s}
+\sum_{s=1}^{\infty}e^{-4\,\eps\,s}\Big(-\frac{3\pi^{2}}{4s}
+\frac{\pi^{2}}{4s}
\Big)\notag \\
&= \frac{1}{2} \pi ^2 \Big( \big[1 +2 \log (1-e^{-2 \eps })\big] \coth \eps -1\Big]  = \frac{1}{\eps}\Big(\pi^{2}\log\eps+\pi^{2}\log 2+\frac{\pi^{2}}{2}\Big)
-\frac{3\pi^{2}}{2}+\mc O(\eps).
\end{align}

\subsection*{$\bullet$ \  (12)(34)(56)}

\begin{figure}[!htb]
\begin{center}
\includegraphics[scale=0.5]{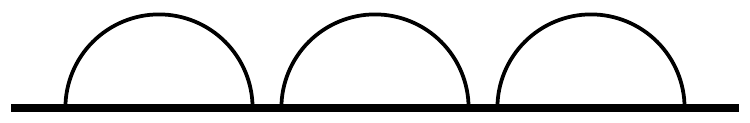}\qquad \qquad
\caption{Ladder diagram associated with the  (12)(34)(56) contraction.
\la{fig:12-34-56} 
}
\end{center}
\end{figure}

Using the same notation,   here  the only non zero case is 
\be
S_{n,m,s} = \begin{cases}
-\frac{\pi^{2}}{2\,s}, & m=n=s, \\ \ \ 
0, & \text{else}.
\end{cases}
\ee
Thus
\begin{align}
I^{(3)}_{(12)(34)(56)} &= -\frac{\pi^{2}}{2}\sum_{s=1}^{\infty}\frac{e^{-3\,\eps\,s}}{s} = 
\frac{1}{2} \pi ^2 \log (1-e^{-3 \eps }) = \frac{\pi^{2}}{2} \, \log \eps +\frac{\pi^{2}}{2} \, \log 3+\mc O(\eps).
\end{align}
The computational  procedure should  now be clear. In the following, we shall just
 present the relevant 
expressions without  detailed comments. %repeating comments. 

\subsection*{$\bullet$ \  (12)(36)(45)}

\begin{figure}[!htb]
\begin{center}
\includegraphics[scale=0.5]{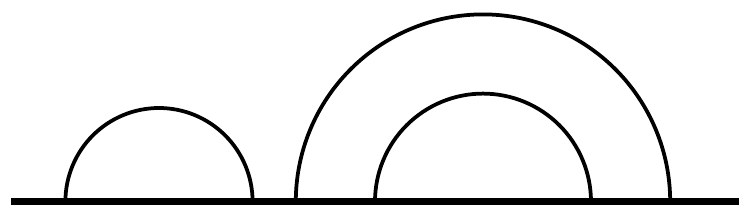}\qquad \qquad
\caption{Ladder diagram associated with the  (12)(36)(45) contraction.
\la{fig:12-36-45} }
\end{center}
\end{figure}

%Here 
\be
S_{n,m,s} = \begin{cases}
\frac{\pi^{2}}{n}, & m=s, \\
0, & \text{else}.
\end{cases}
\ee
\begin{align}
I^{(3)}_{(12)(36)(45)}(\eps) &= \pi^{2}\sum_{n,m=1}^{\infty}\frac{e^{-\eps\,(n+2m)}}{n} = 
-\frac{\pi ^2 \log \left(1-e^{-\eps }\right)}{e^{2 \eps }-1} \no \\
 &= -\frac{\pi^{2}\log\eps}{2\,\eps}+\frac{\pi^{2}}{2}\log\eps+\frac{\pi^{2}}{4}+\mc O(\eps).
\la{142356}
\end{align}

\subsection*{$\bullet$ \  (14)(23)(56)}
\be
S_{n,m,s} = \begin{cases}
\frac{\pi^{2}}{s}, & m=n, \\
0, & \text{else}.
\end{cases}
\ee 
\begin{align}
I^{(3)}_{(14)(23)(56)}(\eps) &= \pi^{2}\sum_{m,s=1}^{\infty}\frac{e^{-\eps\,(s+2m)}}{s} = 
-\frac{\pi ^2 \log \left(1-e^{-\eps }\right)}{e^{2 \eps }-1}\notag \\
&= -\frac{\pi^{2}\log\eps}{2\,\eps}+\frac{\pi^{2}}{2}\log\eps+\frac{\pi^{2}}{4}+\mc O(\eps).
\end{align}
This is equal to the (12)(36)(45)  expression \rf{142356}. Indeed,  the two 
diagrams are   mirror images   under the reflection around $\tau=\pi$, and the measure in the integral  in \rf{2.3} 
  is also  invariant under  this reflection.
%eflection around that point.

\subsection*{$\bullet$ \   (16)(23)(45)}

\begin{figure}[!htb]
\begin{center}
\includegraphics[scale=0.4]{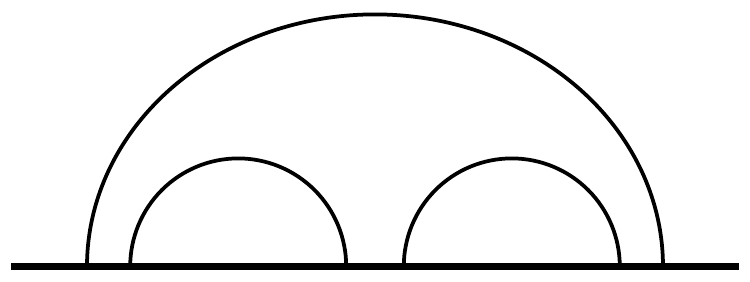}\qquad \qquad
\caption{Ladder diagram associated with the  (16)(23)(45) contraction.
\la{fig:16-23-45} }
\end{center}
\end{figure}

\be
S_{n,m,s} = \begin{cases}
\frac{\pi^{2}\,s}{s^{2}-n^{2}}, & m=n\neq s, \\
\frac{\pi^{2}\,m}{m^{2}-s^{2}}, & s=n\neq m, \\
\frac{\pi^{2}}{s}, & n=m=s,\\
0, & \text{else}.
\end{cases}
\ee
%Thus, 
\begin{align}
I^{(3)}_{(16)(23)(45)}(\eps) &=
\pi^{2}\sum_{s=1}^{\infty}\frac{e^{-3\,\eps\,s}}{s}+
2\, \pi^{2}\mathop{\sum_{m,n=1}}_{m\neq n}^{\infty}\frac{e^{-\eps\,(m+2n)}\,m}{m^{2}-n^{2}} \notag \\
%&=  \pi^{2}\sum_{s=1}^{\infty}\frac{e^{-3\,\eps\,s}}{s}+
% \pi^{2}\mathop{\sum_{m,n=1}}_{m\neq n}^{\infty} e^{-\eps\,(m+2n)}\Big(
% \frac{1}{m-n}+\frac{1}{m+n}\Big) \notag \\
 &=  \frac{\pi^{2}}{2}\,\sum_{s=1}^{\infty}\frac{e^{-3\,\eps\,s}}{s}
 +\pi^{2}\mathop{\sum_{m,n=1}}_{m\neq n}^{\infty} \frac{e^{-\eps\,(m+2n)}}
 {m-n} +
 \pi^{2}\sum_{m,n=1}^{\infty} \frac{e^{-\eps\,(m+2n)}}{m+n} \notag \\
 &= \frac{\pi ^2}{2 (e^{3 \eps }-1)}
\Big[ -(e^{3 \eps }-1) \log (1-e^{-3 \eps
   })+2 (e^{3 \eps }+1) \log (1-e^{-2 \eps
   })\notag \\
   &\qquad + 4 e^{\eps } (e^{\eps }+1) \coth ^{-1}(2
   e^{\eps }+1)-4 \log ( 1 -   e^{-\eps} ) 
   %\sinh (\eps )-\cosh (\eps)+1\big)
   \Big]\notag \\
&= \frac{4\pi^{2}\log 2}{3\eps}+\frac{\pi^{2}}{2}\log \eps
-\frac{\pi^{2}}{2}\log 3-\frac{2\pi^{2}}{3}+\mc O(\eps).
\end{align}

\subsection*{$\bullet$ \   Total  three-loop scalar ladder   contribution:}

\be\la{A.18} 
I^{(3)}_{\text{total}}(\eps) = \frac{\pi^{2}}{\eps}\,\Big(\frac{1}{2}+\frac{7}{3}\,\log 2\Big)
+2\,\pi^{2}\,\log\eps-\frac{5\,\pi^{2}}{3}+\mc O(\eps).
\ee

\subsection{Full $\z$-dependent  two-loop and  three-loop ladder  diagram  contributions} % dependence}
\la{appA3}

To determine the   terms with subleading  powers of  $\z$    we need also to include contributions  with vector exchanges 
in \rf{2.1},\rf{2.2}. 
This is   straightforward to do  using the relation \rf{2.6}   which allows one   to reduce the 
result  to a combination of  lower-loop scalar ladder contributions. 
%\be(\zeta^{2}-\cos\tau)\,\mathscr D(\tau) = (\zeta^{2}-1)\,\mathscr D(\tau)+\tfrac{1}{2}.\ee
%This means that we can use, 
For example,  at two  loops  we have 
\begin{align}\la{A.19} 
& (\zeta^{2}-\cos\tau)\,\mathscr D(\tau)\, (\zeta^{2}-\cos\tau')\,\mathscr D(\tau') = 
\Big[ (\zeta^{2}-1)\,\mathscr D(\tau)+\tfrac{1}{2}\Big]
\Big[ (\zeta^{2}-1)\,\mathscr D(\tau')+\tfrac{1}{2}\Big]\notag \\
&= (\zeta^{2}-1)^{2}\,\mathscr D(\tau)\,\mathscr D(\tau')+\tfrac{1}{2}\,(\zeta^{2}-1)\,
[\mathscr D(\tau)+\mathscr D(\tau')]+\tfrac{1}{4}\ . 
\end{align}
The first term is equivalent to  the one  computed in the previous section  while the term  linear in 
$\zeta^{2}-1$ is proportional to  the  one loop scalar ladder. 

Below we  will   present   the results for the   full  two-loop   and three-loop ladder  contributions
using the notation introduced in \rf{2.7},\rf{2.9}.

\subsection*{ $\bullet$ \     (14)(23)}

The expression for the scalar ladder $I^{(2)}_{(14)(23)} $ was  found already in \rf{A.4}. 
The contribution  $I^{(1)}_{(14)(23)}$  linear in $\z^2-1$    vanishes and thus 
dropping power divergences we get  for the  total  two-loop term in  \rf{2.7},\rf{2.9}  \ci{Beccaria:2017rbe}
\be
I_{(14)(23)} = \frac{\pi^{2}}{2}\,(\zeta^{2}-1)^{2}+\frac{\pi^{4}}{6} \ .
\ee

\subsection*{$\bullet$ \    (12)(34)}

  For $I^{(2)}_{(12)(34)} $  we find 
\begin{align}
S_{n,m} &= n m \int[d^{4}\btau ]
\cos(n\,\tau_{12}) \, \cos(m\,\tau_{34}) = \frac{m^2 (1-\cos 2 \pi  n)- n^2 ( 1 -  \cos 2 \pi  m) }
{m\,n\,(m^{2}-n^{2})}.
\end{align}
For generic integer $n,m$ one checks that $S_{n,m}=0$. Considering separately the 
removable singularity in $n=m$, we find   that  $S_{n,n}=0$, {\em i.e.} 
$I^{(2)}_{(12)(34)}=0$.
For  $I^{(1)}_{(12)(34)}$ we get  from \rf{A.19} a single    sum 
\begin{align}
S_{n} =  \int[d^{4}\btau]\ \frac{1}{2}\,\Big[
-n\,\cos(n\,\tau_{12})-n\,\cos(n\,\tau_{34})\Big]= \frac{1}{n^3}-\frac{\cos (2 \pi  n)}{n^3}-\frac{2 \pi ^2}{n}.
\end{align}
There are no special limits to be considered  and, replacing $\cos(2\pi n)\to 1$,  we get 
\be
I_{(12)(34)}^{(1)} = - 2 \pi^2  \sum_{n=1}^{\infty}e^{-n\,\eps}  \frac{1}{n} 
= 2 \pi ^2 \log (1 - e^{-\eps} ) = 2\,\pi^{2}\,\log\eps + \mc O(\eps).
\ee
As a result, 
\be
I_{(12)(34)} = 2\,\pi^{2}\,(\zeta^{2}-1)\,\log\eps+\frac{\pi^{4}}{6},
\ee
which is same as  the expression in eq.(B.23) in \cite{Beccaria:2017rbe}.

\subsection*{$\bullet$ \    (16)(25)(34)}

Here   we need to add to  the  scalar ladder  $I^{(3)}_{(16)(25)(34)}$ found in \rf{A.9} also  the contributions with two   propagators
\be
I^{(2)}_{(16)(25)(34)}(\eps) = \sum_{n,m=1}^{\infty}e^{-\eps\,(n+m)}\ S_{n,m},
\ \ \ \qquad 
S_{n,m} = \begin{cases}
-\frac{\pi^{4}}{3}+\frac{2\pi^{2}}{n^{2}}, & m=n, \\
0, & \text{else}.
\end{cases}
\ee
Doing the sum gives 
\begin{align}
I^{(2)}_{(16)(25)(34)}(\eps) = 
2 \pi ^2 \text{Li}_2(e^{-2 \eps })-\frac{1}{6} \pi ^4 (\coth \eps -1) = -\frac{\pi ^4}{6\, \eps }+\frac{\pi ^4}{2}+\mc O(\eps)
\end{align}
  $I^{(1)}_{(16)(25)(34)}$  turns out to vanish.

\subsection*{$\bullet$ \    (12)(34)(56)}

$I^{(2)}_{(12)(34)(56)}$ has 
\be
S_{n,m} =\frac{3\pi^{2}}{n\,m} \ , 
\ee
and thus  after summation 
\begin{align}
I^{(2)}_{(12)(34)(56)}(\eps) &= 3 \pi ^2 \log ^2(\sinh \eps -\cosh \eps +1) \notag \\
&= 3 \pi ^2 \log ^2\eps +\mc O(\eps).
\end{align}
We also find 
\begin{align}
I^{(1)}_{(12)(34)(56)}(\eps) &= \sum_{n=1}^{\infty}e^{-n\,\eps}\Big(
\frac{3 \pi ^2}{2 n^3}-\frac{\pi ^4}{2 n}
\Big) \notag \\
&= \frac{1}{2} \Big[3 \pi ^2 \text{Li}_3(e^{-\eps })+\pi ^4 \log (\sinh \
\eps -\cosh \eps +1)\Big]\notag \\
&= \frac{3}{2} \pi ^2 \zetaR (3)+\frac{1}{2} \pi ^4 \log \eps +\mc O(\eps).
\end{align}

\subsection*{$\bullet$ \    (12)(36)(45) = (14)(23)(56)}

Here in $I^{(2)}$ we have 
\be
\la{A.30}
S_{n,m} = \begin{cases}
\frac{2 \pi ^2 (m^3-2 m n^2)}{n (m^2-n^2)^2}, & m\neq n, \\
\frac{5 \pi ^2}{8 n^2}-\frac{\pi ^4}{6}, & m=n.
\end{cases}
\ee
To sum this, we symmetrize the $m\neq n$ case and replace
\begin{align}
\frac{2 \pi ^2 (m^3-2 m n^2)}{n (m^2-n^2)^2} &\to 
\frac{\pi ^2 (m^4-4 m^2 n^2+n^4)}{m n (m^2-n^2)^2}= -\frac{\pi ^2}{2 (n-m)^2}+\frac{\pi ^2}{2 (m+n)^2}+\frac{\pi ^2}{m n}.
\end{align}
The last two terms  are computed by  summing over unconstrained $n,m$
and subtracting the $n=m$ part. The first term  is twice the $n>m$ contribution and can be
found by  setting $n=m+k$ and summing over unconstrained $k,m$. Finally,
we add the $m=n$ case of (\ref{A.30}). The result is
%\footnote{We checked it numerically against the 
%sum at finite $\eps$ of (\ref{A.30}).}
\begin{align}
I^{(2)}_{(12)(36)(45)}(\eps) &= -\frac{1}{12} \pi ^2 \Big[ 6 \text{Li}_2(e^{-2 \eps })+6 
\text{Li}_2(e^{-\eps }) \coth \eps +\pi ^2 (\coth \eps -1)\no \\ 
&\qquad\qquad   -12 \log^2(1- e^{-\eps} )+6 \log (1- e^{\eps} )\Big] \notag \\
&= -\frac{\pi ^4}{6 \eps }+\pi ^2 \log ^2\eps -\pi ^2 \log \eps+\frac{\pi ^2}{2}
+\mc O(\eps).
\end{align}
We  also  find 
\begin{align}
I^{(1)}_{(12)(36)(45)}(\eps) &= \sum_{n=1}^{\infty}e^{-n\,\eps}\Big(
\frac{2 \pi ^2}{n^3}-\frac{\pi ^4}{3 n}
\Big) = 2 \pi ^2 \text{Li}_3(e^{-\eps })+\frac{1}{3} \pi ^4 \log (1 - e^{-\eps} ) \notag \\
&= \frac{1}{3} \pi ^4 \log \eps  + 2 \pi ^2 \zetaR (3)
+\mc O(\eps).
\end{align}

\subsection*{$\bullet$ \    (16)(23)(45)}

Here, we have for $I^{(2)}$ 
\be
\la{A.34}
S_{n,m} = \begin{cases}
-\frac{\pi^{2}}{n\,m}, & m\neq n, \\
0, & \text{else}.
\end{cases}
\ee
%Summing
\begin{align}
I^{(2)}_{(16)(23)(45)}(\eps) &= 
\pi ^2 \Big[  \text{Li}_2(e^{-2 \eps })-\log ^2(1- e^{-\eps})\Big] =  \frac{\pi ^4}{6}-\pi ^2 \log ^2\eps +\mc O(\eps).
\end{align}
$I^{(1)}$ is given by 
\begin{align}
I^{(1)}_{(16)(23)(45)}(\eps) &= \sum_{n=1}^{\infty}e^{-n\,\eps}\Big(
-\frac{\pi ^2}{2 n^3}-\frac{\pi ^4}{6 n}
\Big) \notag \\
&= \frac{ \pi ^4  }{6} \log (1-e^{-\eps} )-{ \pi ^2 \ov 2} 
\text{Li}_3(e^{-\eps })
=
\frac{1}{6} \pi ^4 \log \eps -\frac{\pi ^2 }{2}\zetaR (3)
+\mc O(\eps).
\end{align}
Summing up the  above results   for different diagrams  we   find 
the full ladder  diagram contribution   in \rf{2.12},\rf{2.13}.

\section{Small $\theta_{0}$ expansion of $\langle W^{(\z_{1},\z_{2})}\rangle$}
\la{sec:theta}

We can write the two-loop expression of $\langle W^{(\z_{1},\z_{2})}\rangle$ in (\ref{3.14})
in terms of $\z$ and $\theta_{0}$  using  (\ref{1.5}) 
and expand in  small $\theta_{0}$. % (equivalently, this is the expansion  
%in small $\z_2$  for fixed $\z_1=
At $\theta^2_0$ order  we get  
\be%gin{align}
\la{B.1}
 \frac{\partial^{2}}{\partial\theta_{0}^{2}}\log \langle W^{(\z_{1}, \z_{2})}\rangle 
\Big|_{\theta_{0}=0}= \frac{\z^{2}}{4}\,\Big[
-\lambda+\lambda^{2}\,\Big(\frac{1}{24}+\frac{(\z^{2}-1)\,
[1+\log(\muDR\,R)]}{4\,\pi^{2}}\Big)+\mc O(\lambda^{3})\Big].\qquad 
\ee%nd{align}
For  $\z=1$ this is just  the small $\theta_{0}$ limit of the 
$\frac{1}{4}$-BPS latitude loop and in this case we know that (\ref{B.1}) is proportional to the 
Bremsstrahlung function \cite{Correa:2012at} \footnote{
%Of course, there is nothing non-trivial in this. 
If $\z=1$ in (\ref{3.14}), we simply have the
$\frac{1}{4}$-BPS latitude result which is same as the  $\frac{1}{2}$-BPS circular loop 
expression  with 
$\lambda\to \lambda\,\cos^2 \theta_0$.
}
\begin{align}
\la{B.2}
\left. \frac{\partial^{2}}{\partial\theta_{0}^{2}}\log \langle W^{(\z_{1}, \z_{2})}\rangle 
\right|_{\substack{\theta_{0}=0\\ \z=1}}&= -\frac{\lambda}{4}+\frac{\lambda^{2}}{96}+\dots
=  -4\pi^{2}\,B(\lambda),\qquad
B(\lambda) = \frac{\sqrt\lambda}{4\pi^{2}}\frac{I_{2}(\sqrt\lambda)}{I_{1}(\sqrt\lambda)} .
\end{align}
As is well known, this relation allows one to determine the two-point function coefficient for the 
conformal  operators 
corresponding to  $\Phi^{1}$ and $\Phi^{2}$   in the  defect CFT$_{1}$.  %To see this, 
Eq. (\ref{B.2}) is equivalent to the following correlator of the scalar fields   restricted to the loop 
%$\tau$ dependent field 
\begin{align}
\la{B.3}
& \int_{0}^{2\pi} d\tau\int_{0}^{2\pi} d\tau'
\llangle \widetilde\Phi(\tau)\widetilde\Phi(\tau')\rrangle = -4\,\pi^{2}\,B(\lambda), \qquad
\widetilde\Phi(\tau) = \cos\tau\,\Phi^{1}(x(\tau))+\sin\tau\,\Phi^{2}(x(\tau)).
\end{align}
Conformal symmetry predicts that for  the   %\underline{protected} {\em transverse} 
scalars  $\Phi^{1}$ and $\Phi^{2}$  that are not coupled to the  loop
for   $\theta_0=0$  (and thus have protected  dimension $\Delta=1$)  %we have
\begin{align}
\llangle \Phi^{a}(\tau_{1})\,\Phi^{b}(\tau_{2})\rrangle = \delta^{ab}\,\frac{C_{0}(\lambda)}{
|2\,\sin\frac{\tau_{12}}{2}|^{2\,\Delta}}, \qquad\ \ \   \Delta\equiv1.
\end{align}
Then  (\ref{B.3}) gives the known result \cite{Correa:2012at}
\be
C_{0}(\lambda) = 2\,B(\lambda) =\frac{\lambda}{8\pi^{2}}-\frac{\lambda^{2}}{192\pi^{2}}
+\mc O(\lambda^{3}) . 
\ee
This  follows from 
%Another interesting expansion is found by taking $\z_{1}=1$ and expanding at small $\z_{2}$.
%This amounts to the $\frac{1}{2}$-BPS loop with  (integrated) insertions 
%of the $\tau$ dependent combination 
%coupling to $\z_{2}$, see (\ref{1.4}). At two-loops there is not explicit dependence on $\overline\mu$ 
%and
%\begin{align}
%\log\langle W^{(1, \z_{2})}\rangle &= \log\Big[1+
%\lambda\Big(\frac{1}{8}-\frac{\z^{2}_{2}}{8}+\mc O(\z_{2}^{3})\Big)
%+\lambda^{2}\,\Big(\frac{1}{192}-\frac{\z_{2}^{2}}{96}+\mc O(\z_{2}^{3})\Big)+\mc O(\lambda^{3})
%\Big] \notag \\
%&= \lambda\Big(\frac{1}{8}-\frac{\z^{2}_{2}}{8}+\mc O(\z_{2}^{3})\Big)
%+\lambda^{2}\,\Big(-\frac{1}{384}+\frac{\z_{2}^{2}}{192}+\mc O(\z_{2}^{3})\Big)+\mc O(\lambda^{3})
%\Big].
%\end{align}
%In our previous 1d CFT interpretation this means 
%\begin{align}
%\la{5.30}
%& \int_{0}^{2\pi} d\tau\int_{0}^{2\pi} d\tau'
%\llangle \widetilde\Phi(\tau)\widetilde\Phi(\tau')\rrangle = -\frac{\lambda}{4}+\frac{\lambda^{2}}{96}+\mc O(\lambda^{3}), \notag \\
%&\widetilde\Phi(\tau) = \cos\tau\,\Phi^{1}(\tau)+\sin\tau\,\Phi^{2}(\tau).
%\end{align}
%For  the protected {\em orthogonal} fields $\Phi^{1}$ and $\Phi^{2}$ we expect/know
%\begin{align}
%\la{5.31}
%& \llangle \Phi^{a}(\tau_{1})\Phi^{b}(\tau_{2})\rrangle = \delta^{ab}\,\frac{C_{0}(\lambda)}{
%|2\,\sin\frac{\tau_{12}}{2}|^{2\,\Delta}}, \qquad a,b=1,2\notag \\
%& \Delta=1, \quad C_{0}(\lambda) = 2\,B(\lambda) = \frac{\sqrt\lambda}{2\,\pi^{2}}
%\frac{I_{2}(\sqrt\lambda)}{I_{1}(\sqrt\lambda)}=\frac{\lambda}{8\pi^{2}}-\frac{\lambda^{2}}{192\pi^{2}}+\mc O(\lambda^{3}).
%\end{align}
\foot{Here we need to keep  generic $\Delta$ first   and  send it  to 1 at the end.}
\begin{align}
 \int_{0}^{2\pi} & d\tau\int_{0}^{2\pi} d\tau' 
\llangle \widetilde\Phi(\tau)\,\widetilde\Phi(\tau')\rrangle  = 
2 \,C_{0}(\lambda)\,\int_{0}^{2\pi} d\tau\int_{0}^{2\pi} d\tau'\frac{\cos\tau\cos\tau'}
{(4\sin^{2}\frac{\tau-\tau'}{2})^{\Delta}} = \notag \\
&= 2\,C_{0}(\lambda)\,\int_{0}^{2\pi} d\tau\int_{0}^{2\pi} d\tau'\frac{\cos(\tau+\tau')\cos\tau'}
{(4\sin^{2}\frac{\tau}{2})^{\Delta}} = 
2\,C_{0}(\lambda)\,\int_{0}^{2\pi} d\tau\int_{0}^{2\pi} d\tau'\frac{\cos\tau\cos^{2}\tau'}
{(4\sin^{2}\frac{\tau}{2})^{\Delta}}  \notag \\
&= {2\,\pi}\,C_{0}(\lambda)\,\int_{0}^{2\pi} d\tau\frac{\cos\tau}
{(4\sin^{2}\frac{\tau}{2})^{\Delta}} = \pi^{3/2}\,C_{0}\,\frac{4^{1-\Delta}\,\Delta\,
\Gamma(\frac{1}{2}-\Delta)}{\Gamma(2-\Delta)}\stackrel{\Delta\to 1}{\to } -2\,\pi^{2}\,C_{0},
\end{align}
and comparing with (\ref{B.3}).

%%%%%%%%%%%%%%%%%%%%%%%%%%%%%%%%%%%%%%%
\section{Three-loop ladder diagram  contribution  to $\langle W^{(\z_{1},\z_{2})}\rangle$}
\la{app:3loop2coupl}

%%%%%%%%%%%%%%%%%%%

Let us present   the full ladder diagram three-loop contribution  to 
 the 
$(\z_{1},\z_{2})$ generalized  circular  loop.
 Its  large $\z_i$ limit was  given
in (\ref{3.15}).  % Here, we report the full contribution. Of course, one must 
  %always remind that the  
%Of course, we  emphasize once more that 
 The  subleading terms in this expression ({\em i.e.} terms with non-maximal powers 
of $\z_i$)  will of course get contributions  also from other non-ladder diagrams
and thus  their  ladder  expressions   will be   incomplete. 
Explicitly, 
\begin{align}
\langle & W^{(\z_{1},\z_{2})}\rangle_{\rm ladder} = 1+\frac{\lambda}{8}\,(1-\z_{2}^{2})\notag \\
&+\lambda^{2}\,\Big[
\frac{1}{192}\,(1-\z_{2}^{2})^{2}+\frac{1}{128\,\pi^{2}}\,(\z_{1}^{2}+\z_{2}^{2}-1)^{2}
{+\frac{1}{32\,\pi^{2}}(1-\z_{2}^{2})(\z_{1}^{2}+\z_{2}^{2}-1)\,\log\eps}
\Big]\notag \\
&+ \lambda^{3}\,\Big[
\frac{1}{9216}\,(1-\z_{2}^{2})^{3}
+\frac{1}{768\,\pi^{2}}\,(1-\z_{2}^{2})\,(\z_{1}^{2}+\z_{2}^{2}-1)^{2}\notag \\
&\qquad 
-\frac{1}{1536\,\pi^{4}}(\z_{1}^{2}+\z_{2}^{2}-1)^{2}(5\z_{1}^{2}+8\z_{2}^{2}-8)
+\frac{5}{512\,\pi^{4}}(1-\z_{2}^{2})^{2}(\z_{1}^{2}+\z_{2}^{2}-1)\,\zetaR(3)\notag \\
&\qquad {
+\frac{1}{256\,\pi^{4}}(\z_{1}^{2}+\z_{2}^{2}-1)^{2}(\z_{1}^{2}+2\z_{2}^{2}-2)\log\eps
+\frac{1}{384\pi^{2}}\,(1-\z_{2}^{2})^{2}(\z_{1}^{2}+\z_{2}^{2}-1)\,\log\eps}\notag \\
&\qquad 
{
+\frac{1}{128\,\pi^{4}}\,(1-\z_{2}^{2})(\z_{1}^{2}+\z_{2}^{2}-1)^{2}\,\log^{2}\eps}
\Big]+\dots.
\end{align}
Assuming the {minimal subtraction} scheme, {\em i.e.}  removing only the 
 logarithms of $\eps$, one 
has the two special cases:  $\z_2=0$ and $\z_1=0$.   In the first case  we get \rf{2.13}  with  divergent terms omitted 
\begin{align}
\langle & W^{(\z_{1} ,0)}\rangle_{\rm ladder} = 1+\frac{\lambda}{8}
+\lambda^{2}\,\Big[
\frac{1}{192}+\frac{1}{128\,\pi^{2}}(1-\z_{1}^{2})^{2}
\Big]\notag \\
&+ \lambda^{3}\,\Big[
\frac{1}{9216}+\frac{1}{768\pi^{2}}(1-\z_{1}^{2})^{2}-\frac{1}{1536\,\pi^{4}}\,(1-\z_{1}^{2})^{2}
(5\z_{1}^{2}-8)+\frac{5}{512\pi^{4}}(1-\z_{1}^{2})\,\zetaR(3)
\Big]+\dots,
\end{align}
where   the  $\z_{1}\to 1$  limit    is the circular loop expression
%  we recover the standard 
%Of course, the limit $\z_{1}\to 1$ gives the correct usual expansion of Bessel function. This had to be so,
(all non-ladder diagrams  are known to cancel for $\z_{1}=1$). 
For $\z_1=0$ we get 
\begin{align}
\langle & W^{(0,  \z_{2})}\rangle_{\rm ladder} = 1+\frac{\lambda}{8}\,(1-\z_{2}^{2})
+\lambda^{2}\,(1-\z_{2}^{2})^{2}\,\Big(
\frac{1}{192}+\frac{1}{128\,\pi^{2}}
\Big)\notag \\
&\qquad \qquad + \lambda^{3}\,\,(1-\z_{2}^{2})^{3}\,\Big[
\frac{1}{9216}+\frac{1}{768\,\pi^{2}}+\frac{1}{192\,\pi^{4}}-\frac{5}{512\,\pi^{4}}\,
\zetaR(3)
\Big]+\dots,
\end{align}
and again for $\z_{2}=1$  we reproduce  the known   exact result $\langle W\rangle =1$  for the $1\ov 4$-supersymmetric  loop of \ci{Zarembo:2002an}.

%%%%%%%%%%%%%%%%%%%%%%%
\section{Winding generalization 
of  deformed $\frac{1}{4}$-BPS WM loop: two-loop order}
\la{app:winding}

We can split the calculation of the two-loop diagrams (12)(34) and (14)(23) 
with winding
separating the scalar  and  vector exchanges and the interference term appearing at two loops.

\subsection{Scalar exchanges}

This is the contribution that comes purely from the coupling $\sim\Phi_{m}n^{m}$.

\subsection*{Diagram (14)(23)}

With the same notation as in App.~\ref{sec:3L}, 
we find for generic integer  $\nu>0$ (we explicitly
symmetrize with respect to the exchange $n\leftrightarrow m$)
\be
S_{n,m} = \begin{cases}
\frac{\pi^{2}}{48}\,(-3+8\,\pi^{2}\nu^{2}), & n=m=\nu \\
-\frac{\pi^{2}\,m^{2}\,(m^{2}+\nu^{2})}{2\,(m^{2}-\nu^{2})^{2}}, & n=m\neq \nu \\
-\frac{\pi^{2}m(m+2\nu)}{4(m+\nu)^{2}}, & n=m+2\nu\\
-\frac{\pi^{2}n(n+2\nu)}{4(n+\nu)^{2}}, & m=n+2\nu\\
\frac{\pi^{2}m(m-2\nu)}{4(m-\nu)^{2}}, & m+n=2\nu, n\neq m\\
0, & \text{else}.
\end{cases}
\ee
The regularized sum is (we directly put  $\eps\to 0$ in finite sums)
\begin{align}
I_{(14)(23)}^{\rm scalar}(\eps) &= \frac{\pi^{2}}{48}\,(-3+8\,\pi^{2}\nu^{2})
+\sum_{\substack{n+m=2\nu\\ n\neq m}}\frac{\pi^{2}m(m-2\nu)}{4(m-\nu)^{2}}\notag \\
&\ \ \  -\sum_{m\neq \nu} e^{-2\,m\,\eps}\,\frac{\pi^{2}\,m^{2}\,(m^{2}+\nu^{2})}{2\,(m^{2}-\nu^{2})^{2}}
-2\,\sum_{m} e^{-2\,(m+\nu)\,\eps}\frac{\pi^{2}m(m+2\nu)}{4(m+\nu)^{2}}\ . 
\end{align}
Let us evaluate the relevant infinite sums
\begin{align}
&  -\sum_{m\neq \nu} e^{-2\,m\,\eps}\,\frac{\pi^{2}\,m^{2}\,(m^{2}+\nu^{2})}{2\,(m^{2}-\nu^{2})^{2}}
=  -\sum_{m=1}^{\nu-1}\frac{\pi^{2}\,m^{2}\,(m^{2}+\nu^{2})}{2\,(m^{2}-\nu^{2})^{2}}
 -\sum_{m=\nu+1}^{\infty} e^{-2\,m\,\eps}\,\frac{\pi^{2}\,m^{2}\,(m^{2}+\nu^{2})}
 {2\,(m^{2}-\nu^{2})^{2}}.
\end{align}
The first term here is  finite and  can be computed in closed form. The second contribution  can be simplified
by expressing  it in terms of the basic sum
\begin{align}
\la{D.4}
S_{1}(\eps) &= \sum_{m=\nu+1}^{\infty}\frac{e^{-2\,m\,\eps}}{(m^{2}-\nu^{2})^{2}} = 
\frac{e^{-2 (\nu +1) \eps }}{4 \nu ^3}
\Big[
\nu\,   \Phi (e^{-2 \eps },2,2 \nu +1)+\nu\,  e^{2 \eps } 
\text{Li}_2(e^{-2 \eps })\notag \\
&+e^{2 (2 \nu +1) \eps } B_{e^{-2 \eps }}(2 \nu \
+1,0)+e^{2 \eps } \log (1-e^{-2 \eps })
\Big].
\end{align}
Here $\Phi(z,s,a) =\sum_{k=0}^\infty { z^k\ov |k+a|^s}$ is the Lerch  transcendent function and 
$B_z(a,b) =\int^z_0 dt\,  t^{a-1} (1-t)^{b-1}$ is  the Beta-function. 
The other remaining sum can be   put into the form %expressed in terms of the Hurwitz-Lerch function
\begin{align}
-2\,\sum_{m} e^{-2\,(m+\nu)\,\eps}\frac{\pi^{2}m(m+2\nu)}{4(m+\nu)^{2}} = 
-\frac{1}{4} \pi ^2 e^{-2 (\nu +1) \eps } \Big[-2 \nu ^2 \Phi (e^{-2 \
\eps },2,\nu +1)+\coth \eps +1\Big].
\end{align}
Collecting everything, we find 
\begin{align}
I_{(14)(23)}^{\rm scalar}(\eps) = 
-\frac{\pi ^2}{2 \eps }+\frac{1}{2} \pi ^2 [2 \nu ^2 \psi ^{(1)}(\nu)+2 \nu -1]+\mc O(\eps).
\end{align}

\subsection*{Diagram (12)(34)}

In this case, the $\tau$ integral gives, for generic (positive integer) $\nu$,
\be
S_{n,m}=
\begin{cases}
\frac{\pi^{2}}{12}\,(3+2\,\pi^{2}\nu^{2}), & n=m=\nu \\
\frac{\pi^{2}\,m\,\nu\,(m^{2}+\nu^{2})}{(m^{2}-\nu^{2})^{2}}, & n=\nu, m\neq \nu \\
\frac{\pi^{2}\,n\,\nu\,(n^{2}+\nu^{2})}{(n^{2}-\nu^{2})^{2}}, & m=\nu, n\neq \nu \\
0, & \text{else}.
\end{cases}
\ee
The regularized sum is (we set again $\eps\to 0$ in finite or convergent sums)
\begin{align}
I_{(12)(34)}^{\rm scalar}(\eps) &= \frac{\pi^{2}}{12}\,(3+2\,\pi^{2}\nu^{2})
 +2\,\sum_{m\neq \nu} e^{-(m+\nu)\,\eps}\,
\frac{\pi^{2}\,m\,\nu\,(m^{2}+\nu^{2})}{(m^{2}-\nu^{2})^{2}}.
\end{align}
The infinite sum can be written in terms of $S_{1}(\eps)$ defined in (\ref{D.4}):
\be
I_{(12)(34)}^{\rm scalar}(\eps) = \frac{1}{2} \pi ^2 \nu  \Big[\pi ^2 \nu -4 \nu\, \psi ^{(1)}(\nu )
-4 \psi^{(0)}(\nu )-4 \log \eps -4 \gamma_{\rm E} \Big]+\mc O(\eps).
\ee

\subsection*{Total two-loop contribution from scalar coupling}

Adding the two diagrams, and including the overall  factor $\frac{\lambda^{2}}{(8\pi^{2})^{2}}$,
we find
\be
\la{D.10}
W_{\rm 2-loop}^{\rm scalar} = {-\frac{\lambda^{2}}{32\,\pi^{2}}\,\nu\,\log\eps}
+\frac{\lambda^{2}}{128\,\pi^{2}}\Big[
\pi ^2 \nu ^2-2 \nu ^2\, \psi ^{(1)}(\nu )-4 \gamma_{\rm E}\,  \nu +2 \nu -4 \nu  \
\psi ^{(0)}(\nu )-1
\Big]
\ee

\subsection{Vector exchanges   and scalar-vector interference contributions}

The vector contribution is independent of $\nu$ and may be found  from the $\z$-independent 
part of (\ref{2.13})
\be
\la{D.11}
W_{\rm 2-loops}^{\rm vector} = \lambda^{2}\,\Big[
\frac{1}{192}+\frac{1}{128\,\pi^{2}}{-\frac{1}{32\,\pi^{2}}\,\log\eps}\Big].
\ee
%\subsection{Scalar-Vector Interference}
The interference term coming from 
\be
\frac{\cos(\nu \tau)-\cos\tau}{4\sin^{2}\frac{\tau}{2}}\, 
\frac{\cos(\nu \tau')-\cos\tau'}{4\sin^{2}\frac{\tau'}{2}},
\ee
may be written  as 
\be
-\frac{\cos(\nu \tau)\cos\tau'+\cos\tau\cos(\nu\tau')}{4\sin^{2}\frac{\tau}{2}\,4\sin^{2}\frac{\tau'}{2}}
= \underbrace{-\frac{\cos(\nu \tau)+\cos(\nu\tau')}
{4\sin^{2}\frac{\tau}{2}\,4\sin^{2}\frac{\tau'}{2}}}_{\rm I}
+\underbrace{
\frac{1}{2}\,\Big[\frac{\cos(\nu\tau)}{4\sin^{2}\frac{\tau}{2}}+
\frac{\cos(\nu\tau')}{4\sin^{2}\frac{\tau'}{2}}\Big]}_{\rm II}.\no 
\ee
The first term is a simplified double integral contribution. The second one has  only one denominator
and involves a single sum only.\footnote{This cumbersome splitting may appear as a minor simplification, but 
we found it to be a useful improvement from a  computational point of view.}

\subsection*{Diagram (14)(23): Interference I}

\be
S_{n,m}=
\begin{cases}
\frac{\pi^{2}(2m^{2}+2m \nu+\nu^{2})}{2m(m+\nu)}, & n=m+\nu\\
\frac{\pi^{2}(2n^{2}+2 n \nu+\nu^{2})}{2 n (n+\nu)}, & m=n+\nu\\
-\frac{\pi^{2}(2 n^{2}-2 n \nu+\nu^{2}}{2 n (n-\nu)}, & n+m=\nu \\
0, & \text{else}.
\end{cases}
\ee
%Thus,
\begin{align}
I_{(14)(23)}^{\rm int-I}(\eps)  &= -\sum_{n+m=\nu}\frac{\pi^{2}(2 n^{2}-2 n \nu+\nu^{2})}{2 n (n-\nu)}
+2\,\sum_{m} e^{-(2m+\nu)\eps}\frac{\pi^{2}(2m^{2}+2m \nu+\nu^{2})}{2m(m+\nu)} \no\\
%After  doing the sum, the small $\eps$ expansion is 
\la{D.16}
&= \frac{\pi ^2}{\eps }+\pi ^2 
\Big[-2 \nu +2 \nu  (\psi ^{(0)}(\nu )+\gamma )+1\Big]+\mc O(\eps).
\end{align}

\subsection*{Diagram (14)(23): Interference II}

\be
S_{n}=
\begin{cases}
-\frac{\pi^{4}}{3}\,\nu, & n=\nu\\
0, & \text{else}.
\end{cases}
\ee
%Hence, 
\be
\la{D.18}
I_{(14)(23)}^{\rm int-II}(\eps)  = -\frac{\pi^4}{3}\,\nu+\mc O(\eps).
\ee

\subsection*{Diagram (12)(34) : Interference I}

\be
S_{n,m}=
\begin{cases}
-2\pi^{2}, & n=m=\nu\\
-\frac{\pi^{2}\nu}{m}, & n=\nu, m\neq \nu\\
-\frac{\pi^{2}\nu}{n}, & m=\nu, n\neq \nu\\
0, & \text{else}.
\end{cases}
\ee
%Thus,
\be
I_{(12)(34)}^{\rm int-I}(\eps)  = -2\pi^{2}-2\sum_{m\neq \nu}
e^{-(m+\nu)\eps}\frac{\pi^{2}\nu}{m}= 2\,\pi^{2}\,\nu\,\log\eps+\mc O(\eps).
%\end{align}
%After summing, the small $\eps$ expansion is 
\la{D.21}
%I_{(12)(34)}^{\rm int-I}(\eps)  
\ee

\subsection*{Diagram (12)(34) :  Interference II}

\be
S_{n}=
\begin{cases}
-\frac{\pi^{2}(3+4\pi^{2}\nu^{2})}{12\nu}, & n=\nu\\
-\frac{2 n \pi^{2}(n^{2}+\nu^{2})}{(n^{2}-\nu^{2})^{2}}, & \text{else}.
\end{cases}
\ee
%Hence, 
\begin{align}
I_{(12)(34)}^{\rm int-II}(\eps)  &= -\frac{\pi^{2}(3+4\pi^{2}\nu^{2})}{12\nu}
-2\,\pi^{2}\,\sum_{n\neq \nu}e^{-n\eps}\
\frac{n (n^{2}+\nu^{2})}{(n^{2}-\nu^{2})^{2}}\no 
\\
\la{D.24}
%I_{(12)(34)}^{\rm int-II}(\eps)  
& = \frac{2}{3} \pi ^2 \Big[-\pi ^2 \nu +3 \nu  \psi ^{(1)}(\nu )+3 \psi 
^{(0)}(\nu )+3 \log (\eps )+3 \gamma_{\rm E} \Big]+\mc O(\eps)\ . 
\end{align}

%%%%%%%%%%%%%%%%%%%%%%%%%%%%%%
\subsection*{Total two-loop contribution}

Adding the  expressions 
\rf{D.10},\rf{D.11},\rf{D.16},\rf{D.18},\rf{D.21}  and \rf{D.24} we find
the total in (\ref{4.13}). This is the contribution to the 
$1\ov 4$-supersymmetric  loop $\z_{1}=0$, $\z_{2}=1$
with winding in $S^{5}$.
%\begin{align}
%W_{\rm 2-loops}(\nu) = \frac{\lambda^{2}}{384\,\pi^{2}}\,\Big[
%6\,(1-\nu)+\pi^{2}\,(2-6\nu+3\nu^{2})+12(\psi(\nu)-\psi(1))+6\,(2-\nu)\,\psi^{(1)}(\nu)
%\Big].
%\end{align}
%Special values for $\nu=1,2,3,\dots$ are 
%\be
%W_{\rm 2-loops}(\nu) = 0, \ \lambda^{2}\,(\tfrac{1}{192}+\tfrac{1}{64\pi^{2}}), \ 
%\lambda^{2}\,(\tfrac{1}{48}+\tfrac{19}{256\,\pi^{2}}), \dots
%\ee
%The large $\nu$ expansion (semiclassical limit ? ) is 
%\be
%W_{\rm 2-loops}(\nu) = \frac{\lambda^{2}}{128}\,\Big[
%\nu ^2-2\,\Big(1+\frac{2}{\pi ^2}\Big) \nu +
%\frac{4\, (\log \nu+\gamma_{\rm E}) }{\pi ^2}+\frac{5}{\pi^2}+\frac{2}{3}+\mc O(1/\nu)\Big].
%\ee
The similar result with $\z_{1}=0$ and generic $\z_{2}, \nu$ is obtained by combining the same
pieces. Indeed, purely scalar contributions will have a $\z_{2}^{4}$ factor while the 
scalar-vector interference will have $\z_{2}^{2}$. This leads to the expression 
(\ref{4.10}).

%%%%%%%%%%%%%%%%%%%%%%%%%%%%%%%%%%
\section{One-loop contribution in  dimensional regularization}
\la{app:dimreg}

As a consistency check of the mode regularization, 
let us derive (\ref{4.5}) in dimensional regularization. For the purposes of 
illustration we specialize to the case $\nu_{1}=1$
and $\nu_{2}=\nu\in\mathbb N$. We start with 
\be
W_{1}^{(\z_{1},\z_{2})}(1, \nu) = \frac{\Gamma(\omega-1)}{16\,\pi^{\omega}}\int_{0}^{2\pi} 
d\tau_{1}d\tau_{2}\ 
\frac{\z_{1}^{2}+\z_{2}^{2}\cos(\nu \tau_{12})-\cos(\tau_{12})}
{(4\sin^{2}\frac{\tau_{12}}{2})^{\omega-1}}\ .
\ee
The result will be finite, so we  may %don't bother
ignore the   replacements $\pi^{\omega}\to \pi^{2}$, {\em etc.} % and so on. 
We first write
\begin{align}
W_{1}^{(\z_{1},\z_{2})}(1, \nu) &= \frac{1}{64\,\pi^{2}}\int_{0}^{2\pi} 
d\tau_{1}d\tau_{2}\Big[
\z_{1}^{2}+\z_{2}^{2}\cos(\nu \tau_{12})-\cos(\tau_{12})\Big]\,
(\sin^{2}\tfrac{\tau_{12}}{2})^{1-\omega}\notag \\ 
&= \frac{1}{8}+\frac{\z_{2}^{2}}{64\,\pi^{2}}\int_{0}^{2\pi} 
d\tau_{1}d\tau_{2} \cos(\nu \tau_{12})\,
(\sin^{2}\tfrac{\tau_{12}}{2})^{1-\omega} .
\end{align}
Then, we use the following Chebyshev polynomial expansion valid for integer $\nu\ge 1$
\begin{align}
\cos(\nu \tau)  &= (-1)^{\nu}\,T_{2\,\nu}(\sin\tfrac{\tau}{2}) \no \\ 
&= 
(-1)^{\nu}\,\Big[\nu\,\sum_{k=1}^{\nu}\frac{(-1)^{k}(2\nu-k-1)!}{k!(2\nu-2k)!}
(4\sin^{2}\tfrac{\tau}{2})^{\nu-k}+2^{2\nu-1}(\sin^{2}\tfrac{\tau}{2})^{\nu}\Big],
\end{align} 
and 
\be
\int_{0}^{2\pi}d\tau_{1}d\tau_{2}\ (\sin^{2}\tfrac{\tau_{12}}{2})^{\alpha} = \frac{4\pi^{3/2}\Gamma
(\alpha+\frac{1}{2})}{\Gamma(1+\alpha)}.
\ee
After evaluating the finite summation over $k$, we find 
\be
W_{1}^{(\z_{1},\z_{2})}(1, \nu)  = \frac{1}{8}-
\frac{4^{\omega-3}\,\Gamma(3-2\omega)\Gamma(\omega+\nu-1)\sin(\pi\omega)}{\pi\,\Gamma
(2+\nu-\omega)}\,\z_{2}^{2}.
\ee
Finally, in the four dimensional $\omega\to 2$ limit, we get 
\be
W_{1}^{(\z_{1},\z_{2})}(1, \nu)  = \frac{1}{8}(1-\nu\,\z_{2}^{2}),
\ee
in agreement with (\ref{4.5}).

%\newpage

\bibliography{BT-Biblio}
\bibliographystyle{JHEP}

\end{document}